\documentclass[longauth]{aa}

\usepackage{color}
\usepackage{graphicx}
\usepackage{rotating}
\usepackage{longtable}
\usepackage{amsmath}
\usepackage{amstext}
\usepackage{amssymb}
\usepackage{natbib}

\def   \araa {{\rm {ARA\&A}}}
\def   \apj {{\rm {ApJ}}}

\def   \apjs {{\rm {ApJS}}}

\def   \aap {{\rm {A\&A}}}
\def   \aapr {{\rm {A\&AR}}}

\def   \mnras {{\rm {MNRAS}}}

\def   \apjl {{\rm {ApJL}}}
\def   \nat {{\rm {Nature}}}

\begin{document}

\title{Chasing discs around O-type (proto)stars} 
\subtitle{ALMA evidence for an SiO disc and disc wind from G17.64+0.16\thanks{The data presented in this article are only available in electronic form
at the CDS via anonymous ftp to cdsarc.u-strasbg.fr (130.79.128.5)
or via http://cdsweb.u-strasbg.fr/cgi-bin/qcat?J/A+A/} }
\titlerunning{The disc and disc wind from G17.64+0.16}

\author{L.~T.~Maud\inst{1}\thanks{E-mail:maud@strw.leidenuniv.nl}  \and R.~Cesaroni\inst{2} \and M.~S.~N.~Kumar\inst{3,4} \and F.~F.~S.~van der Tak\inst{5,6}  \and V.~Allen\inst{5,6} \and M.~G.~Hoare\inst{7}
  \and P.~D.~Klaassen\inst{8} \and D.~Harsono\inst{1} \and M.~R.~Hogerheijde\inst{1,9} \and {\'A}.~S{\'a}nchez-Monge\inst{10} \and P.~Schilke\inst{10} \and A.~Ahmadi\inst{11} \and M.~T.~Beltr{\'a}n\inst{2}
  \and H.~Beuther\inst{11} \and T.~Csengeri\inst{12} \and S.~Etoka\inst{13} \and G.~Fuller\inst{13,14} \and R.~Galv{\'a}n-Madrid\inst{15} \and C.~Goddi\inst{1,16}
  \and Th.~Henning\inst{11} \and K.~G.~Johnston\inst{7}  \and R.~Kuiper\inst{17} \and S.~Lumsden\inst{7} \and L.~~Moscadelli\inst{2} \and J.~C.~Mottram\inst{11} \and T.~Peters\inst{18}
  \and V.~M.~Rivilla\inst{2}  \and L.~Testi\inst{19,2} \and S.~Vig\inst{20} \and W.~J.~de Wit\inst{21} \and H.~Zinnecker\inst{22,23}
  }

  \authorrunning{L.~T.~Maud et al.}
  
  \institute{Leiden Observatory, Leiden University, PO Box 9513, 2300 RA Leiden, The Netherlands \email{maud@strw.leidenuniv.nl}\label{inst1}
    \and INAF, Osservatorio Astrofisico di Arcetri, Largo E. Fermi 5, 50125, Firenze, Italy\label{inst2}
    \and Instituto de Astrof\'{i}sica e Ci\^{e}ncias do Espa'c{c}o, Universidade do Porto, CAUP, Rua das Estrelas, 4150–762 Porto, Portugal\label{inst3}
    \and Centre for Astrophysics, University of Hertfordshire, Hatfield, AL10 9AB, UK\label{inst4}
    \and Kapteyn Astronomical Institute, University of Groningen, The Netherlands\label{inst5}
    \and SRON, Landleven 12, 9747 AD, Groningen, The Netherlands\label{inst6}
    \and School of Physics and Astronomy, University of Leeds, Leeds, LS2 9JT, UK\label{inst7}
    \and UK Astronomy Technology Centre, Royal Observatory Edinburgh, Blackford Hill, Edinburgh, EH9 3HJ, UK\label{inst8}
    \and Anton Pannekoek Institute for Astronomy, University of Amsterdam, Science Park 904, 1098 XH, Amsterdam, The Netherlands\label{inst9}
    \and I. Physikalisches Institut der Universit{\"a}t zu K{\"o}ln, Z{\"u}lpicher Str. 77, 50937 K{\"o}ln, Germany\label{inst10}
    \and Max Planck Institute for Astronomy, K{\"o}nigstuhl 17, 69117, Heidelberg, Germany\label{inst11}
    \and Max-Planck-Institut f{\"u}r Radioastronomie, Auf dem H{\"u}gel 69,  53121 Bonn, Germany\label{inst12}
    \and Jodrell Bank Centre for Astrophysics, School of Physics and Astronomy, The University of Manchester, Oxford Road, Manchester M13 9PL, UK\label{inst13}
    \and UK ALMA Regional Centre Node, The University of Manchester, Manchester M13 9PL, UK\label{inst14}
    \and Universidad Nacional Aut{\'o}noma de M{\'e}xico, Instituto de Radioastronom{\'i}a y Astrof{\'i}sica, Morelia, Michoac{\'a}n, 58089, Mexico\label{inst15}
    \and Department of Astrophysics/IMAPP, Radboud University, PO Box 9010, 6500 GL Nijmegen, The Netherlands\label{inst16}
    \and Institute of Astronomy and Astrophysics, University of T{\"u}bingen, Auf der Morgenstelle 10, 72076 T{\"u}bingen, Germany\label{inst17}
    \and Max-Planck-Institut f{\"u}r Astrophysik, Karl-Schwarzschild-Str.1, 85748 Garching, Germany\label{inst18}
    \and European Southern Observatory, Karl-Schwarzschild-Str. 2, 85748   Garching bei M{\"u}nchen, Germany\label{inst19}
    \and Indian Institute of Space science and Technology, 695 547,  Thiruvananthapuram, India\label{inst20}
    \and European Southern Observatory, Al{\'o}nso de Cordova 3107, Vitacura, Casilla, 19001, Santiago de Chile, Chile\label{inst21}
    \and Deutsches SOFIA Institut, Pfaffenwaldring 29, Univ. Stuttgart, D-70569 Stuttgart, Germany\label{inst22}
  \and Universidad Autonoma de Chile, Av. Pedro Valdivia 425, Santiago de Chile, Chile\label{inst23}}

  \date{Received 19 July 2018 / Accepted 13 September 2018}

\abstract{We present high angular resolution ($\sim$0.2$''$) continuum and molecular emission line Atacama Large Millimeter/sub-millimeter Array (ALMA) observations of G17.64+0.16 in Band 6 (220$-$230\,GHz) taken as part of a campaign in search of circumstellar discs around (proto)-O-stars. At a resolution of $\sim$400\,au the main continuum core is essentially unresolved and isolated from other strong and compact emission peaks. We detect SiO (5-4) emission that is marginally resolved and elongated in a direction perpendicular to the large-scale outflow seen in the $^{13}$CO (2$-$1) line using the main ALMA array in conjunction with the Atacama Compact Array (ACA). Morphologically, the SiO appears to represent a disc-like structure. Using parametric models we show that the position-velocity profile of the SiO is consistent with the Keplerian rotation of a disc around an object between 10$-$30\,M$_{\odot}$ in mass, only if there is also radial expansion from a separate structure. The radial motion component can be interpreted as a disc wind from the disc surface. Models with a central stellar object mass between 20 and 30\,M$_{\odot}$ are the most consistent with the stellar luminosity (1$\times$10$^5$\,L$_{\odot}$) and indicative of an O-type star. The H30$\alpha$ millimetre recombination line (231.9\,GHz) is also detected, but spatially unresolved, and is indicative of a very compact, hot, ionised region co-spatial with the dust continuum core. The broad line-width of the H30$\alpha$ emission (full-width-half-maximum = 81.9\,km\,s$^{-1}$) is not dominated by pressure-broadening but is consistent with underlying bulk motions. These velocities match those required for shocks to release silicon from dust grains into the gas phase. CH$_3$CN and CH$_3$OH thermal emission also shows two arc shaped plumes that curve away from the disc plane. Their coincidence with OH maser emission suggests that they could trace the inner working surfaces of a wide-angle wind driven by G17.64 which impacts the diffuse remnant natal cloud before being redirected into the large-scale outflow direction. Accounting for all observables, we suggest that G17.64 is consistent with a O-type young stellar object in the final stages of protostellar assembly, driving a wind, but that has not yet developed into a compact H{\sc ii} region. The existance and detection of the disc in G17.64 is likely related to its isolated and possibly more evolved nature, traits which may underpin discs in similar sources.}


\maketitle

\section{Introduction}
\label{intro}

The formation scenario for the most massive stars ($>$8\,M$_{\odot}$) still remains uncertain. From a theoretical stand point models generally invoke a scaled-up solar-mass model, with a disc, or disc-like structure, accretion streams and axial jets or outflows \citep[e.g.][]{Krumholz2009,Peters2010,Kuiper2010,Kuiper2011,Klassen2016,Rosen2016,Tanaka2017,Kuiper2018} as otherwise massive young stellar objects (YSOs) could not accrete in the face of intense stellar winds and radiation pressure \citep[e.g.][]{Nakano1989, Jijina1996,Wolfire1986}. Observations indicate that jets and outflows are associated with massive YSOs (e.g. \citealt{Beuther2002,Maud2015,Purser2016,Bally2016}), however observations of the putative discs in the mm-regime are scarce, as somewhat expected considering that pre-ALMA, mm-interferometers did not typically resolve targets below $\sim$1000\,au, except in a few studies \citep[e.g.][]{Jimenez2012,Wang2012,Maud2013b,Beuther2013,Cesaroni2014,Hunter2014}. \citet{Beltran2016} provide a comprehensive review of discs around luminous YSOs, in which B-type sources are indicated to have characteristics of scaled-up solar-type systems. Based on previous studies the authors reported that young O-type stellar sources (L$>$10$^{5}$\,L$_{\odot}$) show large-scale rotation in $\sim$10000\,au, potentially transient, toroidal `pseudo-disc' structures \citep[see also][]{Furuya2008}, although relatively recent ALMA observations by \citet{Johnston2015} did indicate a Keplerian signature in a 2000\,au radius structure around a $\sim$25\,M$_{\odot}$ O-type source. Keplerian-like signatures are also found around B-type binary systems, due to a rotating circumbinary disc \citep[e.g.][]{SanchezMonge2013,SanchezMonge2014,Beltran2016a}, while O-type star forming regions show much more complex structures at millimetre wavelengths with anisotropic accretion and spiral-like feeding filaments from large parsec scales to small $\sim$1000\,au scales (e.g. \citealt{Liu2015}, \citealt{Maud2017}, \citealt{Izquierdo2018}, \citealt{Goddi2018}) akin to the streams seen in the aforementioned models. Critically, the latter examples do not show evidence for Keplerian-like discs, at least down to the 500\,au scales that these authors probe. 

Any massive stellar source from $\sim$8\,M$_{\odot}$ can burn hot enough to completely ionise the surrounding molecular material to form an H{\sc ii} region \citep{Wood1989,Churchwell1990,Kurtz2005} and destroy any complex chemical tracers that are traditionally used to understand the kinematics of their natal environments and their accretion discs. Initial arguments by \citet{Walmsley1995} argued that high accretion rates will cause a very dense H{\sc ii} region that is optically thick to radio emission and thus the H{\sc ii} would not be seen. However, for heavily accreting massive YSOs the onset of the H{\sc ii} region could be delayed via stellar bloating \citep{Palla1992,Hosokawa2009,Hosokawa2010,Kuiper2013} where the effective temperature of the star is much cooler than it would be considering a main sequence star of the same luminosity. In these models, a halt or considerable reduction in accretion ($\ll$10$^{-3}$\,M$_{\odot}$\,yr$^{-1}$), or growth beyond $\sim$30$-$40\,M$_{\odot}$, will result in the YSO contracting to a `main-sequence' configuration, heating significantly, and being able to create an H{\sc ii} region. The fine details of this transition are still somewhat unclear since they depend on the assumed accretion law and the initial conditions chosen for the stellar evolution calculations \citep{Haemmerle2016}. Observations in the infra-red (IR) have been made in search of cool stellar atmospheres that point to bloated stars, however these studies remain inconclusive \citep{Linz2009,Testi2010}. Alternative scenarios are that H{\sc ii} regions can be gravitationally trapped at very early ionisation stages \citep[see][]{Keto2003,Keto2007}, or flicker due to chaotic shielding of the ionising radiation by an accretion flow, leading to a non-monotonous expansion \citep{Peters2010b}. Hyper Compact (HC) H{\sc ii} regions (<0.03\,pc) are thought to be the earliest ionisation stage and therefore could relate to the halt of accretion, be a marker of a transition phase.

Radio observations provide some evidence of the transition stages. Insofar, three high-mass YSOs, S106 IR \citep{Hoare1994}, S140 IRS1 \citep{Hoare2006, Maud2013b} and G298.2620$+$00.7394 (as a candidate from \citealt{Purser2016}) are potentially at a stage contracting down from bloated stars and now driving ionised disc winds. These objects have compact, but elongated, radio continuum emission inconsistent with a jet scenario. The compact radio emission for these sources can be understood as an equatorially driven radiative disc wind \citep[e.g.][]{Drew1998}, where the intense radiation from the now contracted YSO is ionising and ablating the disc surface and driving a wind in the equatorial plane (see also \citealt{Lugo2004,Kee2016,Kee2018,Kee2018b}). For S106 IR, the outflow cavity is also ionised, resulting in the striking bi-polar H{\sc ii} region \citep{Schneider2018} and pointing to a more `evolved' stage compared to S140 IRS1, where only the disc is ionised \citep{Hoare2006}.  

Other high-mass sources, MWC349A \citep{Baez2014,Zhang2017} and Orion Source I \citep{Matthews2010,Greenhill2013,Issaoun2017,Ginsburg2018} also drive disc winds. There is still some debate as to whether MWC349A is a pre- or post-main sequence star, while in Orion Source I the radio emission is not consistent with an equatorial ionised disc wind \citep{Matthews2010}. In these cases, the disc winds from these sources co-rotate with the disc (see also \citealt{Klaassen2013}). For MWC349A, the rotation is seen in maser emission \citep{Zhang2017} as was also the first clear case of rotation in Orion Source I via SiO maser analysis \citep{Goddi2009,Matthews2010} before it was recently found to be rotating in thermal emission as well \citep{Hirota2017}. Co-rotating disc winds, are different from the aforementioned equatorial wind scenario where the winds are radiatively driven, as they instead are thought to be intrinsically linked with the accretion process and disc \citep[e.g.][]{Turner2014}. Identifying these disc winds in massive YSOs could therefore provide insight on accretion and mass loss processes.

In this article we report on our target G17.64+0.16 (hereafter G17.64, also known as AFGL 2136, G017.6380+00.1566, CRL 2136, and IRAS 18196-1331). We originally targeted G17.64 along with five other luminous O-type (proto)stars\footnote{For consistency with \citet{Cesaroni2017} we use the same nomenclature `(proto)star' only here. Throughout the paper we use young-stellar-object (YSO) to avoid any ambiguities as massive sources ($>$10\,M$_{\odot}$) are probably buring deuterium and those at later stages beginning to burn hydrogen \citep[e.g.][]{Hosokawa2010} and are not really true `proto'-stars.} with ALMA and found mixed evidence for discs around the various targets using the typical CH$_3$CN tracer \citep{Cesaroni2017}. Our detailed investigations to-date of two of these six sources, G31.41+0.31 and G24.78+0.08 \citep{Beltran2018, Moscadelli2018}, found clear velocity gradients in the CH$_3$CN 12-11 line transitions. In G31.41+0.31, the gradient is indicative of rotational spin-up and infall towards the two central cores, whereas for G24.78+0.08, it is due to wind and outflow feedback from the O-star at the centre of an HCH{\sc ii} region. G17.64, presented in this work, is thought to be a massive O-type YSO at a later formation stage, but prior to forming a compact H{\sc ii} region.

G17.64 is located at 2.2\,kpc and has a bolometric luminosity of 1$\times$10$^5$\,L$_{\odot}$. As part of the Red MSX survey\footnote{http://rms.leeds.ac.uk/cgi-bin/public/RMS\_DATABASE.cgi} \citep{Lumsden2013} G17.64 has the most up-to-date luminosity estimated using the detailed multi-wavelength spectral energy distribution method from \citet{Mottram2011} updated to include fluxes from the \textit{Herschel} HiGAL survey \citep{Molinari2010}, and combined with an unambiguous kinematic source distance \citep{Urquhart2012,Urquhart2014}. The luminosity positions G17.64 as an O9$-$O9.5 type based on \citet{Vacca1996} with an expected mass of 20\,M$_{\odot}$, hence being selected as one of our six sources to study O-star discs \citep{Cesaroni2017}. G17.64 is bright at near- and mid-IR wavelengths (2.2 to 24.5\,$\mu$m $-$ \citealt{Kastner1992,Holbrook1998,deWit2009,Murakawa2013}) as a reflection nebula that is co-spatial with the cavity walls carved by the blue-shifted lobe of the large-scale molecular outflow at a position angle of $\sim$135$^{\circ}$ \citep{Kastner1994,Maud2015} and perpendicular to a suggested polarisation disc \citep{Murakawa2008}. \citet{Murakawa2013} analysed the IR spectrum from G17.64 and noted that the weak Br$\gamma$ line is extemely broad (FWHM = 133\,km\,s$^{-1}$), indicative of an underlying wind. The surrounding molecular cloud has a mass of $\sim$600\,M$_{\odot}$ within a $\sim$1.2$\times$1.2\,pc region \citep{Maud2015a}. G17.64 also has a plethora of other observations ranging from the IR to radio \citep[e.g.][]{Minchin1991,vandertak2000a,vandertak2000b,Menten2004,Wang2007,Lu2014} that image the bi-polar nebulae, the compact radio emission and surrounding environment in a range of typical molecular species such as NH$_3$, CS, CH$_3$OH at mm wavelengths. Notably, G17.64 was also observed using mid-IR interferometry \citep{dewit2011,Boley2013}. Models to reproduce the single MIDI baseline emission presented by \citet{dewit2011} suggest G17.64 could have a compact disc. In the sample of massive YSOs investigated by \citet{Boley2013} the combination of Keck data and multi-baseline MIDI observations also suggest a compact ($<$100\,au diameter) dust disc is present in G17.64. 

In Sect.\,\ref{obs} we present the observation overview, while the main observational results are shown in Sect.\,\ref{res}. In Sect.\,\ref{holder1} we analyse the data using a parametric model and also discuss G17.64 in reference to other known massive YSOs. Finally the conclusions are given in Sect.\,\ref{conc}.

\section{Observations}
\label{obs}
Our ALMA 12\,m observations were conducted during Cycle 2 in July and September 2015 (2013.1.00489.S - PI: Cesaroni). The spectral setup included one wide-band spectral window (SPW) with $\sim$1.875\,GHz bandwidth in order to obtain a continuum estimate and numerous other SPWs with higher spectral resolution and narrower bandwidths $\sim$234\,MHz for the molecular line emission in Band 6 (220$-$230\,GHz). The individual datasets had between 38 and 41 antennas at any one time and baselines ranging from 40\,m out to $\sim$1500\,m. We refer the reader to \citet{Cesaroni2017} for more details on the setup.

The data reduction was undertaken using {\sc casa} version 4.5.3 \citep{McMullin2007}. Owing to G17.64 having weaker molecular line emission than the other O-stars in the sample \citep[c.f.][]{Cesaroni2017}
there were many line-free regions in all SPWs and thus a standard continuum subtraction was undertaken followed by self-calibration using the line-free continuum emission\footnote{The {\sc statcont} package by \citealt{SanchezMonge2018} was used for continuum subtraction in the other targets of \citet{Cesaroni2017} which have very few line-free channels. Comparisons were made for G17.64 using {\sc statcont} which did produce comparable products although continuum subtraction in the image plane does not allow for self-calibration after the analysis.}. The self-calibration solutions were applied to the entire G17.64 data and the continuum subtraction was remade, thus both the continuum and line data have the self-calibration applied. The high signal to noise of G17.64 during each iterative stage of self-calibration allowed us to self-calibrate down to a minimal timescale of 6\,seconds (i.e. starting with the source scan time of approximately six minutes, and subsequent iterations using 180, 60 and 30\,seconds, then finally down to the integration time of 6\,s). The improvement of the continuum signal-to-noise was almost a factor of five, improving from $\sim$180 to over $\sim$860. 

A Briggs robust weighting of 0.5 \citep{Briggs1995} was used for all images to balance resolution and sensitivity, while a robust value of 1.5 was also used for the continuum in order to weight towards shorter baselines, sensitive to larger scale diffuse emission. Table \ref{table1} presents the beam sizes, noise levels and velocity resolution of the various images used in this work. We note that a subset of these data were first presented in \citet{Cesaroni2017}, namely images of SiO and CH$_3$CN, which were analysed in a systematic fashion along with the other target sources. Here however our data have undergone re-imaging and analysis after self-calibration. The self-calibration notably improved the signal-to-noise of our images and given the significant improvement in phase correction, have acted to sharpen the image structures and boost the detected fluxes \citep[see also][]{Cornwell1999}. Our results superseed those presented in \citet{Cesaroni2017}.

\begin{table*}[ht!]
\begin{center}
  \caption{Noise, beam size and position angle, and velocity resolution for the 12\,m array continuum and line data images made in this work. A robust weighting of 0.5 was used unless indicated. For molecular lines the transition and upper energy levels are also listed. Units are as indicated in the column heading unless otherwise stated.}
  {\footnotesize
\begin{tabular}{@{}lrrrcrr@{}}
\hline
\hline
Image  &  Frequency & Transition  & E$_u$   & Beam   &  Noise$^{\rm a}$  &    Resolution  \\
      &   (GHz)     &           &  (K)      &  ($''$,$^{\circ}$)    &(mJy/beam)  &  (km\,s$^{-1}$) \\
\hline
Continuum  & 226.700 &   ...     & ...   &   0.17$\times$0.14, $-$88.2  &  0.07   &         2.65 (GHz)              \\
Continuum (robust 1.5) & 226.700  & ...  &   ...  &  0.21$\times$0.17, $-$87.9  &  0.16   &         2.65 (GHz)          \\
\hline
$^{13}$CO   &  220.39868  &  2$-$1  & 15.87 & 0.18$\times$0.15, $+$89.8   &  $<$4.03$^{\rm b}$   &   0.35  \\
SiO        &  217.10498  &  5$-$4  &  31.26  & 0.20$\times$0.15, $-$88.3   &  1.21    & 2.70  \\
CH$_3$CN  &  220.70902 & 12$-$11 K=3 & 133.16  &  0.18$\times$0.15, $-$89.2   & 2.08  &   0.35  \\
CH$_3$OH       &  218.44005 & 4$_{(2,2)}-3_{(1,2)}$-E  &  45.46     & 0.20$\times$0.15, $-$88.4   & 1.16   & 2.70  \\
H30$\alpha$  &  231.90093 & recomb. line  & ...  &  0.17$\times$0.15, $-$90.0    &  1.58     &  1.40    \\
\hline
\label{table1}

\end{tabular}
  }
\end{center}
{\footnotesize
  $^{\rm a}$ The noise is per channel for line data cubes. \\ $^{\rm b}$ The $^{13}$CO has resolved out structure leading to higher noise in channels where there is strong, poorly sampled emission. Line free channels have a noise of 1.76\,mJy\,beam$^{-1}$\,ch$^{-1}$.}
\end{table*}

A stand-alone ACA and Total Power (TP) project, specifically aimed at understanding the outflows from our selected targets, observed G17.64 as part of Cycle 4 on 27 November 2016 (2016.1.00288.S - PI: Cesaroni). For the ACA data 11 antennas were available and the array had baselines ranging from 8.9 to 45.0\,m. The total time on source was about three minutes for the ACA. Because of this disproportionately short on-source time, the ACA data have a very sparse \emph{u,v} coverage and little overlap with the 12\,m baselines, and are of a low sensitivity. Therefore the ACA data are unsuitable for merging with the aforementioned 12\,m data in creating a representative image or well formed and weighted synthesised beam. The total power observations had $\sim$19\,minutes on-source time and used data from three antennas in single dish mode. The TP pointing were set to cover the primary beam area of the ACA only $\sim$45$''\times$45$''$.

The spectral setup was selected to cover the same transitions as the previous 12\,m array configuration and hence covered SiO and $^{13}$CO, which are usually associated with outflow emission. The data were pipeline reduced and then the ACA data were manually imaged using {\sc casa} version 4.7.2 (more recent data reduction uses a newer {\sc casa} version). The final beam size was $\sim$7.38$''\times$5.04$''$ with a position angle of 84.6$^{\circ}$ when using a robust weighting of 0.5. The noise ranges from 0.2 to $\sim$1.0\,Jy\,beam$^{-1}$ per 0.5\,km\,s$^{-1}$ channel (the latter from channels with significant emission). For the TP data we use the delivered ALMA products. We note that the outflow parameters from these ACA data and a comparison to the TP data are presented in Appendix \ref{AppendixA} and are not presented in the main paper.

\section{Results} 
\label{res}
In this paper we focus on the continuum and the kinematics of the SiO, while also presenting the emission from CH$_3$CN, CH$_3$OH and H30$\alpha$ in order to describe the putative disc and evidence for a disc wind. We also make a reference to the $^{13}$CO observations from the 12\,m array only, in order to describe the orientation of the molecular outflow `bubble'.

\subsection{Continuum emission}
\label{contemm}
\begin{figure}
\begin{center}
\includegraphics[width=0.52\textwidth]{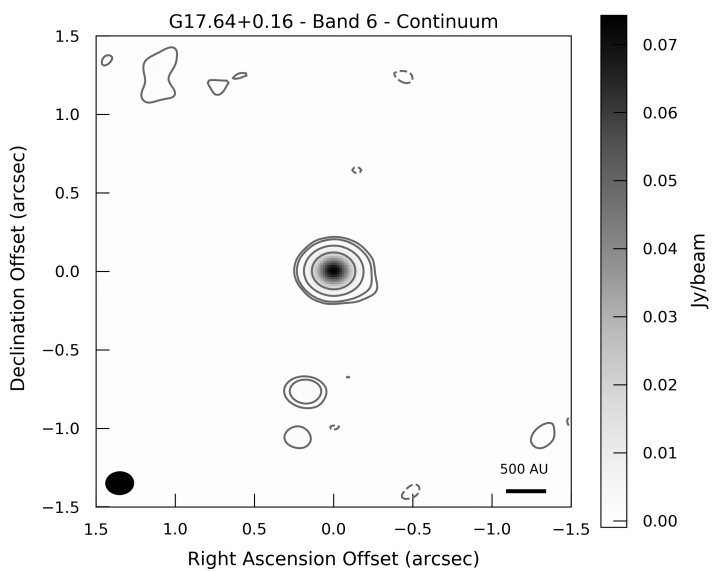}
\caption{ALMA band 6 (1.3\,mm) continuum image of G17.64 made with a robust 0.5 value. The contour levels are set at -3, 5, 10, 50 and 200\,$\sigma$  (1\,$\sigma$ = 0.07\,mJy\,beam$^{-1}$) to indicate the weak southern sources and the strength of G17.64 (negative contours are dashed lines). The map is centred (0,0) on G17.64 at J2000 18$^{\rm h}$22$^{\rm m}$26.385$^{\rm s}$ $-$13$^{\circ}$30$'$11.97$''$. The synthesised beam is indicated to the bottom left of the figure, while a scale bar is shown in the bottom right.} 
\label{fig:fig1} 
\end{center}
\end{figure}

Figure \ref{fig:fig1} shows the 1.3\,mm dust continuum map of G17.64 imaged in our ALMA observations within a 3$''\times$3$''$ region centred on G17.64 (J2000 18$^{\rm h}$22$^{\rm m}$26.385$^{\rm s}$ $-$13$^{\circ}$30$'$11.97$''$). Due to the high dynamic range, the contours at the 5 and 10\,$\sigma$ levels best highlight the very weakest emission in the region. G17.64 is a very compact, bright source (peak = 74.2\,mJy\,beam$^{-1}$, flux density = 81.3\,mJy) and is essentially isolated from other strong and compact cores. The strongest central emission component associated with G17.64 remains unresolved. The next strongest source in the field is a comparatively very weak (peak = 1.8\,mJy\,beam$^{-1}$) point source located 0.7$''$ ($\sim$1540\,au) to the south east, with another point-source further south and even weaker (peak = 0.7\,mJy\,beam$^{-1}$). As we achieve such a high sensitivity in the continuum map, we also detect diffuse continuum emission at the 5 to 10\,$\sigma$ level and a number of point sources throughout the map, which are more clearly seen in Fig.\,\ref{fig:fig2}. 

The central source, G17.64, is associated with compact radio emission \citep{Menten2004} that would contribute a maximum of 29.5\,mJy (36\,\% of the 1.3\,mm flux density) when scaled to the 1.3\,mm wavelength using the free-free spectral index of 1.2 estimated by \citet{Menten2004} from frequencies $<$43.2\,GHz. These authors also argue that the free-free spectral index will probably become shallower beyond 43.2\,GHz, in which case if we assume a $-$0.1 optically thin free-free spectral index the contribution of the free-free at 1.3\,mm would only be 3.57\,mJy (4\,\% of the 1.3\,mm flux density). Thus, in the case where we assume optically thin millimetre emission and consider typical parameters for dust opacity coefficient, where $\kappa_0$ = 1.0 cm$^2$\,g$^{-1}$ as suggested for densities of 10$^6-$10$^8$\,cm$^{-3}$ for dust with thin ice mantles \citep{Ossenkopf1994}, a gas-to-dust ratio of 100, and temperatures between 50 and 100\,K, the mass of the compact dust emission ranges from 0.85 to 2.69\,M$_{\odot}$ when considering the extremes of the free-free contribution (see also Eq.\,1 of \citealt{Maud2013a}). The dust mass should be interpreted with caution as we resolve out emission with the interferometer, we do not correct for dust opacity, and we use a single common temperature. Considering these points together, uncertainties in dust mass by a factor of 2-3 are not unreasonable.

A two-dimensional gaussian fit to the central continuum emission associated with G17.64 in the image plane results in deconvolved source dimensions (FWHM - full-width-half-maximum) of 53 $\times$ 39 milliarcsecond (117\,au $\times$ 86\,au) at a position angle (PA) of 25$\pm$5$^{\circ}$ (east of north). When considering the orientation of the IR-reflection nebula and the large-scale $^{13}$CO molecular outflow (see also Fig.\,\ref{fig:fig2} and Appendix \ref{AppendixA}) at a PA of $\sim$135$^{\circ}$, the deconvolved PA of the dust structure could suggest that the mm emission is tracing an underlying disc in an almost perpendicular direction. The result is also consistent with the IR-interferometric observations by \citet{Boley2013} where their gaussian fitting suggested a disc FWHM of 44 milliarcsecond (97\,au at 2.2\,kpc) with a PA around 38$^{\circ}$. Although we have very high signal-to-noise, we note that the strong centrally compact dust emission is essentially unresolved and therefore warn that the deconvolved parameters should be interpreted with some caution. Table \ref{table2} indicates the continuum parameters for G17.64.

\begin{table*}[ht!]
\begin{center}
  \caption{Continuum observation parameters for G17.64+0.16.}
  {\footnotesize
\begin{tabular}{@{}lcccccc@{}}
\hline
\hline
Name &  Coordinates &  Wavelength  &  Peak Flux  & Flux Density  & Free-free flux  &  Mass$^a$  \\
    &    (J2000)    &   (mm)       &  (mJy\,beam$^{-1}$)  & (mJy)  &  (mJy)     &  (M$_{\odot}$)  \\
\hline
G17.64+0.16  &  18$^{\rm h}$22$^{\rm m}$26.385$^{\rm s}$ $-$13$^{\circ}$30$'$11.97$''$  & 1.3  &  74.2&  81.3  & 3.57$-$29.5  & 0.85$-$2.69 \\
\hline
\label{table2}

\end{tabular}

  }
\end{center}
{\footnotesize
  $^{\rm a}$ Mass extremes estimated using dust temperatures between 50 and 100\,K, a gas-to-dust ratio of 100, and a dust opacity coefficient of 1.0\,cm$^2$\,g$^{-1}$}
\end{table*}

\subsection{Larger scales and the $^{13}$CO molecular outflow}
\label{coemm}
\begin{figure*}[!ht]
\begin{center}
\includegraphics[width=0.90\textwidth]{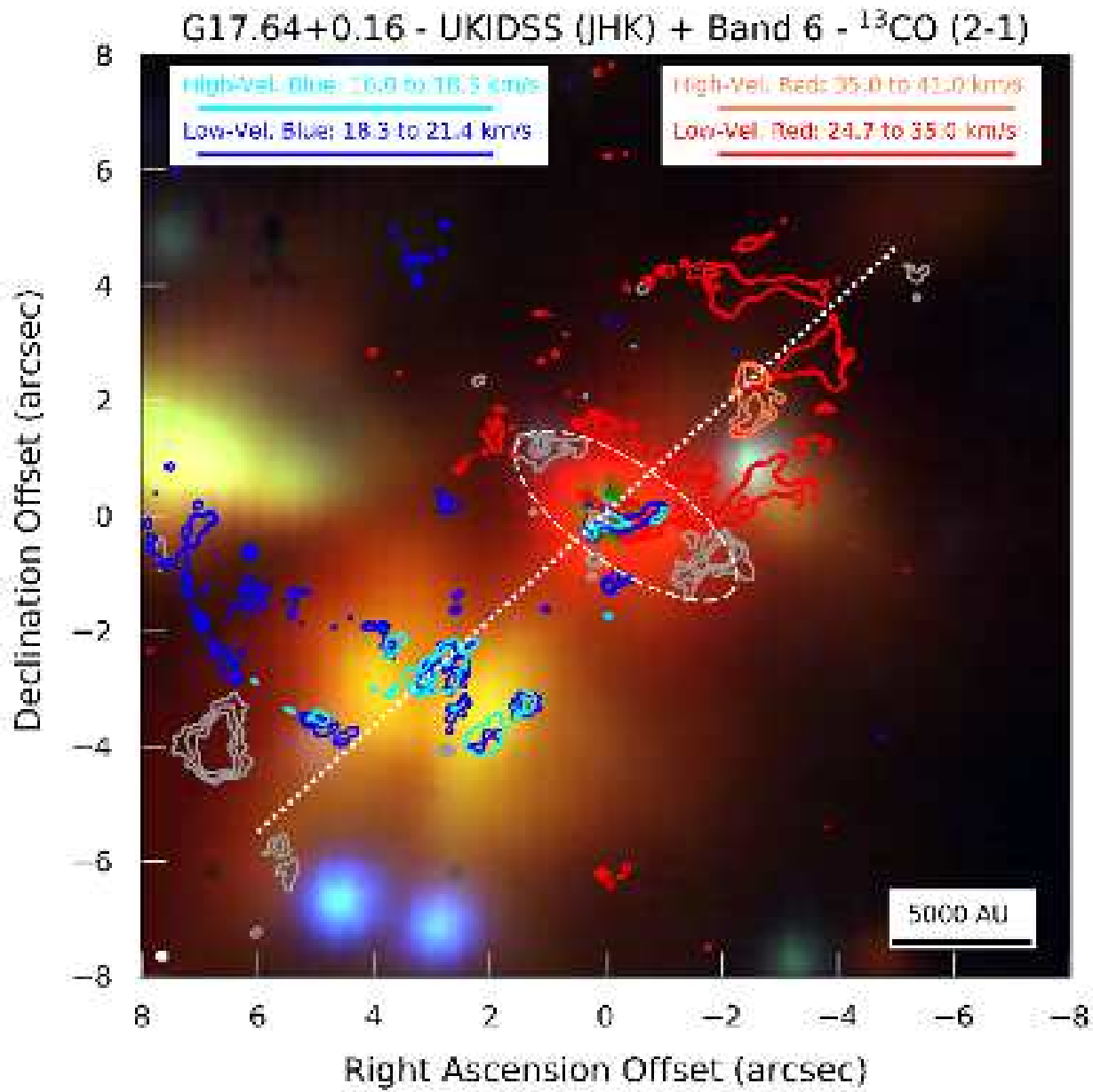}   
\caption{Three colour near-IR composite image made using Ks (2.2\,$\mu$m, red), H(1.6\,$\mu$m, green) and J(1.2\,$\mu$m, blue) images obtained by the UKIDSS overlaid with grey contours of the ALMA band 6 (1.3\,mm) continuum emission (robust 1.5 weighting) from the 12\,m array only. The contours are at the 3, 4, 5\,$\sigma$ levels (1\,$\sigma$ = 0.16\,mJy\,beam$^{-1}$) to highlight the weaker emission features. The position of peak continuum emission from G17.64 is indicated by the plus symbol at (0,0) offset (J2000 18$^{\rm h}$22$^{\rm m}$26.385$^{\rm s}$ $-$13$^{\circ}$30$'$11.97$''$) - the continuum contours are not plotted due to the overlap with $^{13}$CO contours. In the IR, G17.64 at the centre is the dominant source, while to the east, west and south-east the emission is associated with the reflection nebula. The white dashed ellipse is indicative of a possible remnant dusty toroid, or underlying dark lane structure. The coloured contours show the moment zero maps of the $^{13}$CO outflow emission integrated over various velocity ranges. The cyan and orange contours show the blue- and red-shifted highest velocity emission integrated from 16.0 to 18.3\,km\,s$^{-1}$ and 35.0 to 41.0\,km\,s$^{-1}$ respectively. The contour levels are at 5, 10 and 15\,$\sigma$, where $\sigma$ is 3.28 and 4.57\,mJy\,beam$^{-1}$\,km\,s$^{-1}$ for the blue- and red-shifted high velocities. The blue and red contours indicate the integration of the lower outflow velocities from 18.3 to 21.4\,km\,s$^{-1}$ for the blue-shifted emission and from 24.7 to 35.0\,km\,s$^{-1}$ for the red-shifted emission. We note that the source V$_{\rm LSR}$=22.1\,km\,s$^{-1}$. These contours are at 4 and 5\,$\sigma$, where $\sigma$ is 6.25 and 10.06\,mJy\,beam$^{-1}$\,km\,s$^{-1}$ for the blue- and red-shifted velocities. The dotted white line indicates the outflow axis at a PA of $\sim$135$^{\circ}$ \citep[][see also Appendix \ref{AppendixA}]{Kastner1994,Maud2015} which passes though G17.64 and the points of highest velocity blue- and red-shifted outflow emission along with the IR nebula. A synthesised beam for the mm emission is shown to the bottom-left, a scale bar to the bottom-right, and the legend for the integrated velocities at the top. The primary beam cut off is beyond the map extent.}
\label{fig:fig2} 
\end{center}
\end{figure*}

A large-scale view of the G17.64 region is shown in Fig.\,\ref{fig:fig2}. The background is a three colour composite image made using Ks (2.2\,$\mu$m, red), H(1.6\,$\mu$m, green) and J(1.2\,$\mu$m, blue) images. These near-IR images were obtained by the UKIDSS (The UKIRT Infrared Deep Sky Survey, \citealt{Lawrence2007}) Galactic Plane Survey \citep{Lucas2008} and have point spread function of $\sim$0.5$-$0.8$''$ respectively \citep[see also][]{Murakawa2008}. This is overlaid with grey contours at the 3, 4 and 5\,$\sigma$ levels indicating the 1.3\,mm continuum ALMA image made with a robust 1.5 weighting. The outflow emission from $^{13}$CO is shown by the coloured contours while the outflow direction is indicated by a white dotted line. G17.64 is identified as a plus symbol located at (0,0) in the map and, for reference, has a V$_{\rm LSR}$ of 22.1\,km\,s$^{-1}$. 
The two sources to the south of G17.64 (as shown in Fig.\,\ref{fig:fig1}) are clearly seen, while there is also another compact source located in the region of blue-shifted $^{13}$CO emission (at a 2.8, -4.0 offset from G17.64), although only at the 4-5\,$\sigma$ level. There is diffuse continuum emission about 1.5 to 2.0$''$ (3300$-$4400\,au) to the north-east and south-west of G17.64, in a direction almost perpendicular to that of the outflow, suggestive of a remnant from a larger scale possibly dusty toroidal structure or filamentary dark lane (dashed white ellipse). Although our observations are sensitive to scales up to $\sim$7$''$ the diffuse emission has a low surface brightness (peak $\sim$5$-$10\,$\sigma$) and appears disjointed, although we do not consider this to be real sub-structure. Far to the south-east ($\sim$6$''$, $\sim$13200\,au) there is also another diffuse structure, again peaking at around the 10\,$\sigma$ level and of low surface brightness, such that we may only image the peak even large-scale diffuse emission. Lower resolution ($\sim$0.8$''$) shorter baseline observations better image the surrounding extended, weak, dust emission and are consistent with our data in that we only detect the peaks of emission while larger structures are present (Avison et al. 2018 in prep.). The diffuse emission we detect is located in the regions devoid of IR emission. All other continuum features in our map, but not mentioned, are at the 3\,$\sigma$ level.   

In our observations the $^{13}$CO emission serves as a tracer for the outflow as we do not cover the more abundant typically used outflow tracer, $^{12}$CO, in our spectral setup. The $^{13}$CO clearly highlights only the strongest emission regions in the outflow as the emission is over-resolved with our 12\,m main array configuration. At the highest velocities, the blue-shifted emission (16.0 to 18.3\,km\,s$^{-1}$) peaks to the south-east while the red-shifted emission (35.0 to 41.0\,km\,s$^{-1}$) peaks to the north-west. This high-velocity emission follows a roughly linear path through G17.64 and is highlighted by the cyan and orange contours respectively in Fig.\,\ref{fig:fig2}. The blue-shifted emission also appears co-spatial with the bright IR features to the south-east. The lower velocities (blue-shifted from 18.3 to 21.4\,km\,s$^{-1}$, red-shifted from 24.7 to 35.0\,km\,s$^{-1}$ (recall the V$_{\rm LSR}$ is 22.1\,km\,s$^{-1}$), represented by the blue and red contours trace the cavities resulting from a bubble-like shape created by the outflow. These appear to visually match to the broadened outflow cavities shown in \citet{Kuiper2018}. The red-shifted emission appears to trace almost a tear-drop shape with an origin centred on G17.64, while the blue-shifted emission is only visible as an extended arc shaped structure far to the south-east of G17.64. The red-shifted bubble appears as roughly symmetric about the outflow axis and confined at the base within the diffuse dust emission either side of G17.64, meanwhile the blue-shifted arc is located inward (towards G17.64) of the diffuse dust emission to the south-east. The origin of the $^{13}$CO emission appears to be from the working surfaces of the outflow interacting with the surrounding diffuse dust in IR dark regions. All estimates of the outflow parameters are made using only the ACA data which are presented in Appendix \ref{AppendixA}.

Notably, there is also overlapping blue- and red-shifted emission located to the west of G17.64 (Fig.\,\ref{fig:fig2}, RA offset $-$0.5$''$). In the context of an outflow this can only be explained if there is a wide cavity opening angle such that the flow close to G17.64 is emitted almost radially in the plane of a disc, and thus has both blue- and red-shifted components close to the star before the flow is redirected to follow the parabolic shape of the cavity where the bulk motion becomes overall red-shifted to the north-west \citep[e.g.][]{Arce2013}. 

\subsection{SiO}  
\label{Ssio}

\begin{figure*}
\begin{center}
\includegraphics[width=0.92\textwidth]{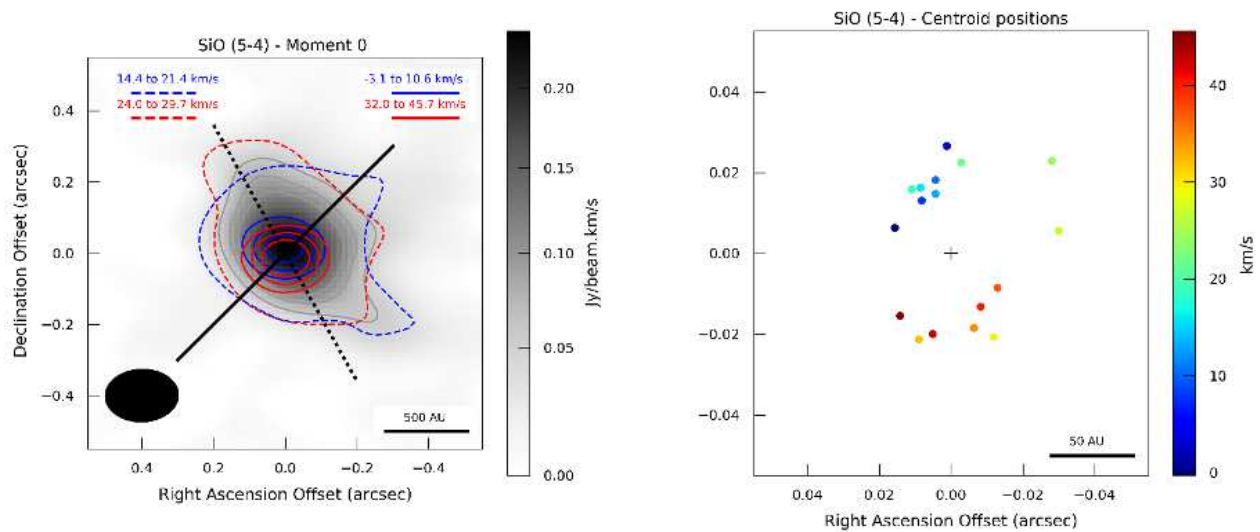}
\caption{Left: Moment zero image of the SiO emission integrated over the V$_{\rm LSR}$ between 21.4 and 24.0\,km\,s$^{-1}$. The grey contours highlight the elongated emission and are at the 10, 15 and 20\,$\sigma$ levels ($\sigma$=5.524\,mJy\,beam$^{-1}$\,km\,s$^{-1}$). The blue and red solid contours show blue- and red-shifted high velocity SiO emission thought to be tracing the high velocity component, possibly an underlying rotating disc (levels at 10, 15 and 20\,$\sigma$, where $\sigma$ = 9.57 and 9.41 mJy\,beam$^{-1}$\,km\,s$^{-1}$ for the blue- and red-shifted high velocity emission) while the dashed lines indicate the 10\,$\sigma$ level ($\sigma$ = 7.54 and 6.58 mJy\,beam$^{-1}$\,km\,s$^{-1}$ for the blue- and red-shifted emission) attributed to the lower velocity component. The velocity ranges are indicated to the top of the figure, a scale bar to the bottom right, and the synthesised beam to the bottom left. The black solid line is the direction of the outflow at a PA of 135$^{\circ}$ and the dotted is that of the disc major axis at a PA of 30$^{\circ}$, corresponding to the cut used for the PV analysis (see Fig.\,\ref{fig:fig4}) Right: Plot of the centroid positions from a 2D Gaussian fitting of each channel from the SiO image cube between -0.3 to 45.6\,km\,s$^{-1}$, clearly showing that the blue- and red-shifted emission are predominantly to the north-east and south-west respectively. We note that the scale is changed (by a factor of ten) compared to the left panel. A scale bar is also indicated to the bottom right.}
\label{fig:fig3} 
\end{center}
\end{figure*}

Figure \ref{fig:fig3} (left) shows the moment zero map of the SiO (5-4, E$_u$ = 31.26\,K) emission integrated over the central $\sim$3\,km\,s$^{-1}$ of the line covering the V$_{\rm LSR}$, between 21.4 and 24.0\,km\,s$^{-1}$. In contrast to \citet{Cesaroni2017}, here we integrate the SiO emission over a narrower velocity range to highlight the elongated structure at near V$_{\rm LSR}$ velocities. The elongated structure of the emission is clearly visible and perpendicular to the large-scale outflow (Fig.\,\ref{fig:fig2}). Fitting the integrated SiO emission with a 2D gaussian in the image plane indicates a PA of $\sim$47$\pm$7$^{\circ}$ and a major axis of $\sim$0.28$''$ ($\sim$600\,au). The deconvolved PA is 30$\pm$10$^{\circ}$ consistent with the PA of the dust continuum emission. The SiO emission is broad in velocity, ranging from -3.0 to 45.7\,km\,s$^{-1}$ considering emission above 3\,$\sigma$, this is broader than presented in \citet{Cesaroni2017}, 4 to 39.1\,km\,s$^{-1}$, owing to our improved images after self calibration. In most massive YSOs, SiO is very spatially extended, typically tracing emission at shock fronts \citep[e.g.][]{Schilke1997} created by active jets and outflows driven by the energetic central sources \citep[e.g.][]{Cabrit2007,Klaassen2015,SanchezMonge2013b,duartecabral2014, Cunningham2016, Cesaroni2017, Beltran2018, Moscadelli2018}. In at least one massive YSO, Orion source I, the SiO emission is rather compact, associated with SiO masers and is shown to trace a rotating disc and disc wind structure \citep{Goddi2009,Ginsburg2018}. The SiO emission from G17.64 is also relatively compact, unlike a jet. There is also compact emission from other silicon and sulphur bearing species in our observations, emission from SiS (v=0 12-11, E$_u$ = 69.95\,K) and $^{33}$SO (6$_5$-5$_4$, E$_u$ = 34.67\,K) are centrally peaked at the location of the continuum and SiO emission. The strong detection of SiO points to a reservoir of silicon in the gas phase, which, in addition to the detected sulphur-bearing species, suggest an association with ionised, hot ($>$100\,K), turbulent and shocked gas \citep[e.g.][]{Minh2010,Minh2016} where these species are released from the grain material.

The solid contours in Fig.\,\ref{fig:fig3} (left) represent the moment zero emission integrated over the highest blue- and red-shifted velocities of the SiO emission (between -3.1 to 10.6\,km\,s$^{-1}$ and 32.0 to 45.7\,km\,s$^{-1}$ respectively) whereas the dashed lines outline the emission integrated over only the low velocity emission (between 14.4 to 21.4\,km\,s$^{-1}$ and 24.0 to 29.7\,km\,s$^{-1}$ for the blue- and red-shifted emission respectively). The PA of a simple line drawn through the centroids of the corresponding blue- and red-shifted emission at the high velocity extreme is consistent with that from the integrated map over the central velocities and with the PA at $\sim$30$^{\circ}$ of the continuum emission. Observationally, the SiO emission from G17.64 appears inconsistent with a jet or outflow origin when considering the known CO outflow and the IR nebulosity at a PA of $\sim$135$^{\circ}$ and the lack of any evidence for a jet with an axis close to the putative disc PA of $\sim$30$^{\circ}$ \citep{Kastner1992,Murakawa2008}. Rather, the SiO appears to originate from a disc-like structure perpendicular to the outflow. Intriguingly, the spatially extended lower velocity blue- and red-shifted emission (dashed contours) appear offset in the opposite direction from the spatially compact high velocity blue- and red-shifted emission (solid contours). If we were to assume a rotating disc is present, the high ($\pm$20$-$25\,km\,s$^{-1}$) and low ($\pm$5$-$10\,km\,s$^{-1}$) velocity rotational directions disagree, that is, with one suggesting a small disc rotating one way - with blue-shifted emission to the north-east and the other indicating a larger disc in the opposite direction - with blue-shifted emission to the south-west (see below and Sect.\,\ref{mods}). In the right panel of Fig.\,\ref{fig:fig3} we show the centroid positions after fitting Gaussians to each channel in the SiO image cube between -0.3 to 45.6\,km\,s$^{-1}$. It is clear that the strongest emission components of each channel are within 0.03$''$ of the central location of G17.64, and moreover that the blue- and red-shifted emission are offset to the north-east and south-west respectively, supporting the idea of a rotating structure. The typical errors on the fits are of the order half a pixel, corresponding to 0.0075$''$. As the strongest emission components from all channels match the conventional rotation of a disc, the extended emission preferentially seen at low velocities and appearing to show opposite rotation due to the spatial extent, probably originates from a kinematically separate structure that we cannot fully disentangle in the observations.

\begin{figure*}[ht!]
\begin{center}
\includegraphics[width=0.97\textwidth]{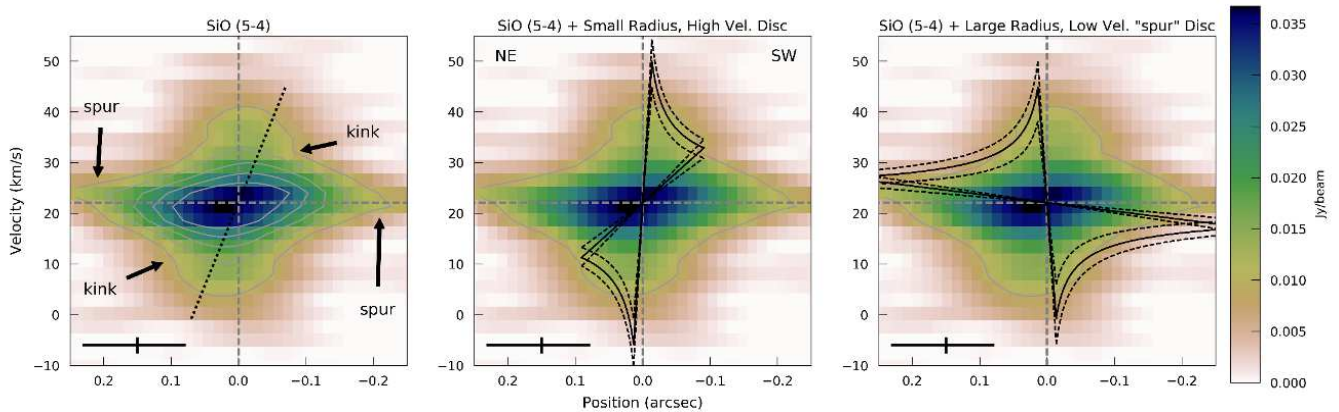}
\caption{Left: Position-velocity (PV) diagram of the SiO emission extracted from a cut (as wide as the synthesised beam) at a position angle of 30$^{\circ}$ across the putative disc major axis (see Fig.\,\ref{fig:fig3} - position offsets are positive to north-east). The grey contours are at the 10, 15, 20, 25 and 30\,mJy\,beam$^{-1}$ level to highlight the structure. The black dotted line is a guide to indicate the excess of high velocity emission at small spatial offsets. Centre: Observed SiO PV as the left plot (colour map with grey contour at a 10\,mJy\,beam$^{-1}$ level) overlaid with pure Keplerian rotation profiles for a 200\,au radius disc around a 20, 30 and 40\,M$_{\odot}$ (inner dashed, solid, outer dashed lines) O-type YSO, when considering an inclination of 70$^{\circ}$ (\citealt{dewit2011}). The disc outer radius was limited to 200\,au such that the lowest velocities roughly matched the kink in the observed data. The disc lines here represent only the high-velocity emission. Right: Observed SiO PV as the left and central plots but overlaid with the contours of pure Keplerian rotation profiles for a disc around a 10, 20, 30\,M$_{\odot}$ (inner dashed, solid, outer dashed lines) OB-type YSO rotating in the opposite direction to that plotted in the central panel (using an inclination of 70$^{\circ}$). The disc outer radius was set to 650\,au in an attempt to match the low velocity spurs at large spatial offsets. The spatial and velocity resolution are indicated by the black bars at the bottom left of all plots. Note the central figure indicates the north-east and south-west offset directions that relate to the cut along the disc direction in Fig.\,\ref{fig:fig3}.}
\label{fig:fig4} 
\end{center}
\end{figure*}

In Fig.\,\ref{fig:fig4} we show the position-velocity (PV) information of the SiO emission extracted from the image cube along a cut at a PA of 30$^{\circ}$, that is, along the major axis direction of the putative disc from the north-east to south-west direction. We include all data along the cut within a width equal to that of the synthesised beam, which helps to slightly improve the signal-to-noise. Overlaid on all images are dashed grey lines to cut the four quadrants. The left image includes grey contours to indicate the general structure of the emission along with a black dotted line to guide the eye as to the skew of the PV diagram indicating high velocities at small offsets ($\pm$0.1$''$). Such a shape in the PV plane can generally be interpreted as an unresolved rotating structure, although we additionally see considerable large-offset emission at low velocities (`spurs') at both blue- and red-shifted velocities (see Sect.\,\ref{mods}). We notice that there appears to be a `kink' in the lowest level grey contour that could be interpreted as a separation between different components, such as rotation (close to linear structure highlighted by the black dotted line) and radial motion, infall or expansion (represented by a diamond like shape, e.g. \citealt{Tobin2012}).  

For illustration we over-plot lines of Keplerian rotation for various central masses \citep[c.f.][]{Cesaroni2014} to explain either the high velocity, small offset emission from a compact disc around a 20, 30 or 40\,M$_{\odot}$ mass O-type YSO, or the lower velocity, large offset emission as a larger disc rotating in the opposite sense about a 10, 20, or 30\,M$_{\odot}$ mass OB-type YSO (central and right panels of Fig.\,\ref{fig:fig4} respectively). We used a source inclination of 70$^{\circ}$, where 90$^{\circ}$ is a view edge-on to the disc, based upon models by \citet{dewit2011}. We do not suggest that there are two counter-rotating discs in G17.64, but present both separate scenarios. Furthermore, we note that our data only marginally resolve the structures in the PV image compared to studies of low-mass YSOs and as predicted for ALMA high-resolution (sub-100\,au) observations from models \citep[e.g.][]{Yen2014,Harsono2015,Seifried2016,Dutrey2017} and thus `fitting' Keplerian rotation lines is not reliable. It is clear that any of the Keplerian rotation curves only represents emission in two of the four quadrants in the PV plots and thus a disc alone cannot fully describe the data, independent of disc size or rotation direction. Furthermore, pure radial motion, infall or expansion, would result in a symmetric diamond shaped PV plot \citep[see][]{Ohashi1997,Brinch2008,Tobin2012,Sanna2018} which alone would not represent the skew seen to high velocities at small offsets. A combination of kinematics components is required. A more detailed analysis, taking also the observational limitations (spatial and velocity resolution) into account is presented in Sect.\,\ref{mods}.

\subsection{CH$_3$CN and CH$_3$OH}
\label{chtracers}
CH$_3$CN is one of the most commonly used tracers to study the kinematics of small scale ($<$0.1\,pc) dense gas surrounding massive YSOs. It is specifically used to investigate the presence of disc signatures and can also probe kinetic temperatures using the various emission lines from the close in frequency `K' rotational ladders \citep[][]{Araya2005,Cesaroni2014, Beltran2014, SanchezMonge2014, Purcell2006}. For G17.64 we detect the J=12-11, K-ladder from K=0 to K=7, E$_u$ = 68.86$-$418.63\,K (although only at the 2-3\,$\sigma$ level for the highest K-lines, K=5$-$7). In correspondence with \citet[][Fig\,6]{Cesaroni2017}, who plot only a relatively cooler K=2 line, we find no evidence for a rotating disc in this molecular tracer, as the CH$_3$CN emission is preferentially detected only to the south-west of the continuum peak of G17.64 (within 1$''$, 2200\,au) throughout all velocities.

\begin{figure*}[ht!]
\begin{center}
  \includegraphics[width=1.0\textwidth]{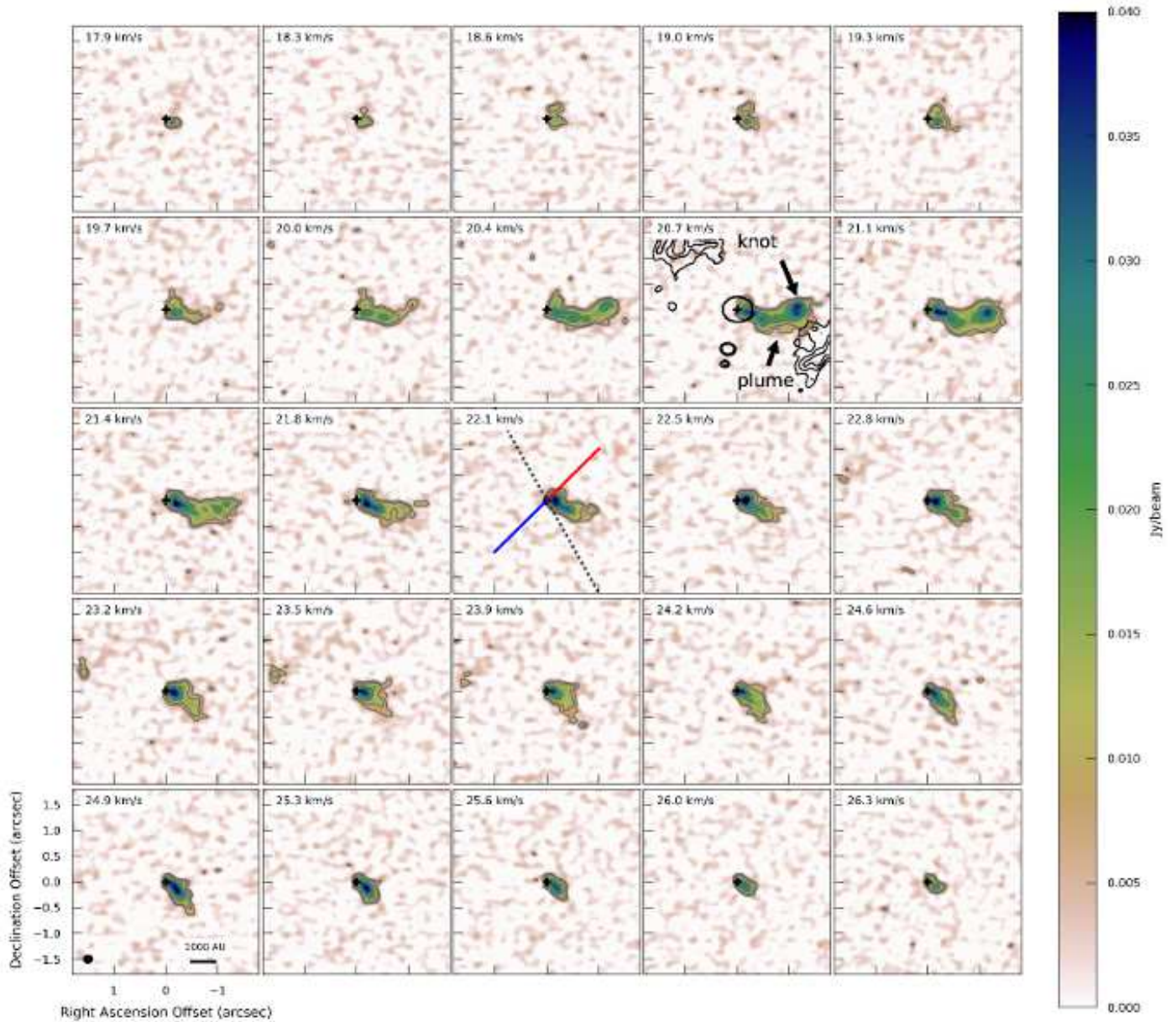}
\caption{Channel map of the CH$_3$CN (J=12-11) K=3 line ranging from 17.9 to 26.3\,km\,s$^{-1}$ in steps of 0.3\,km\,s$^{-1}$. The contours of the CH$_{3}$CN emission are at the 3, 5 and 10\,$\sigma$ level ($\sigma$=2.0\,mJy\,beam$^{-1}$\,chan$^{-1}$). In the central panel at the V$_{\rm LSR}$ velocity channel (22.1\,km\,s$^{-1}$) the dotted line indicates the PA of the disc major axis at 30$^{\circ}$, while the solid (blue and red) line indicates the outflow direction at a PA of 135$^{\circ}$. At all velocities the CH$_{3}$CN emission is brightest to the south west of the main continuum peak (at 0,0 indicated by the '$+$' symbol). The plume structure (indicated) is visible between $\sim$19.7 to $\sim$22.5\,km\,s$^{-1}$ while the knot at the tip of the plume is inward of the diffuse dust emission, potentially indicative of an interaction of outflowing material with the surrouding dust structures. The continuum emission at the 3, 4, 5\,$\sigma$ levels as in Fig.\,\ref{fig:fig2} is plotted in black contours in the 20.7\,km\,s$^{-1}$ channel panel to highlight the diffuse dust emission. A scale bar and synthesised beam are shown in the bottom left panel.}
\label{fig:fig5} 
\end{center}
\end{figure*}

Figure \ref{fig:fig5} shows the channel map from the CH$_3$CN (J=12-11) K=3 line (E$_u$ = 133.16\,K) within a 3.0$\times$3.0$''$ region centred on the dust continuum peak associated with G17.64. A plume shaped structure is evident at marginally blue-shifted velocities (19.7$-$22.5\,km\,s$^{-1}$). At these velocities all low-K transitions (K=0$-$4, E$_u$ = 68.86$-$183.15\,K) show the curved `plume'-like shape. A `knot' feature at the tip of the plume coincides with the inner edge of the diffuse continuum dust emission 1-2$''$ (2200$-$4400\,au) south-west of G17.64, possibly an emission enhancement due to interaction with the surrounding material, the dusty toroid or dark lane structure (see the continuum contours in the 20.7\,km\,s$^{-1}$ panel, Fig.\,\ref{fig:fig5}). Emission from the plume is co-spatial with the $^{13}$CO (see Fig.\,\ref{fig:fig2}), potentially from a wide-angle flow driven by G17.64, and supports such an interpretation. Indeed, if a flow was following a cavity with a wide opening angle, we would also expect some spatial overlap of the  blue- and red-shifted emission close to the source. Furthermore, the blue-shifted emission from K$<$3 lines has a distinctive arc of emission to the south-east that is at the leading edge of the blue-shifted arc already identified in $^{13}$CO, from the interaction of the larger scale outflow bubble with the surrounding medium (this is not shown in Fig.\,\ref{fig:fig5} but the CH$_3$CN matches the $^{13}$CO arc structure in Fig.\,\ref{fig:fig2} to the south-east).

\begin{figure*}
\begin{center}
  \includegraphics[width=0.94\textwidth]{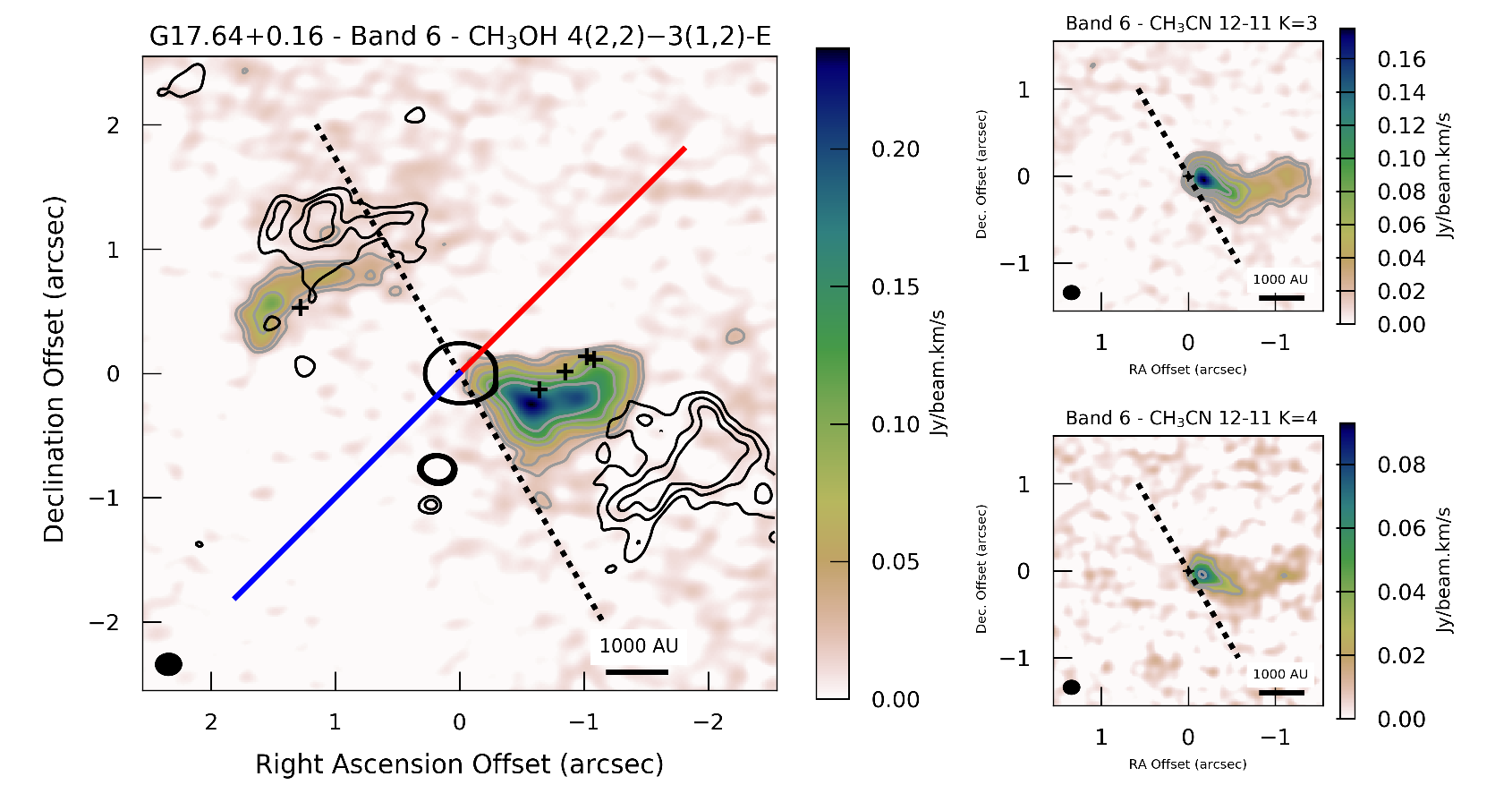}
\caption{Left: Moment zero map of the CH$_3$OH line at 218.440\,GHz integrated between 19.0 and 23.9\,km\,s$^{-1}$ (colour map). The contours of the CH$_{3}$OH emission (grey) are at the 3, 5, 10, 15 and 20\,$\sigma$ level ($\sigma$=7.44\,mJy\,beam$^{-1}$\,km\,s$^{-1}$), while those of the continuum emission at the 3, 4 and 5\,$\sigma$ levels are also shown in black (see also Fig.\,\ref{fig:fig2}). The plume emission both to the east and west curves away from the plane of the proposed SiO disc plane (black dotted line) possibly due to a redirection or collimating of a wide-angle flow to the common CO outflow direction (blue and red solid line). The plus symbols mark the OH masers from \citet{Argon2000}. G17.64 is centred at (0,0). Right: Moment zero maps of CH$_3$CN (J=12-11) K=3 (top) and K=4 (bottom) also integrated between 19.0 and 23.9\,km\,s$^{-1}$ (colour map). The contours are at the 3,5,10,15,20\,$\sigma$ levels ($\sigma$ = 4.93\,mJy\,beam$^{-1}$\,km\,s$^{-1}$ for the K=3 transition and 7.61\,mJy\,beam$^{-1}$\,km\,s$^{-1}$ for the K=4 transition. The dotted line in all panels is that of the SiO disc plane.}
\label{fig:fig6} 
\end{center}
\end{figure*}

In Fig.\,\ref{fig:fig6}(left) we present the moment zero map of the CH$_3$OH line at 218.440\,GHz (4$_{(2,2)}-3_{(1,2)}$-E, E$_u$ = 45.46\,K) integrated between 19.0 and 23.9\,km\,s$^{-1}$ to highlight the plume structures curving away from the plane of the assumed dust continuum and SiO disc (PA $\sim$30$^{\circ}$  - faint dotted line). Both plumes are coincident with the over-plotted OH masers detected by \citet{Argon2000}, not only spatially, but also in velocity. The north-east OH maser is at 23.3\,km\,s$^{-1}$ while those to the south-west are between 19.9 and 20.9\,km\,s$^{-1}$. Multiple other CH$_3$OH (including $^{13}$CH$_{3}$OH) lines are detected in the region surrounding G17.64, but all are weaker. In the right panels of Fig.\,\ref{fig:fig6} we also show the moment zero maps of CH$_3$CN (J=12-11) K=3 (E$_u$ = 133.16\,K) and K=4 (E$_u$ = 183.15\,K) integrated over the same range as CH$_3$OH (between 19.0 and 23.9\,km\,s$^{-1}$), which also indicate the plume structure. The K=4 emission is notably weaker than the K=3 line of CH$_3$CN.

The west plume like structure is mostly associated with blue-shifted emission as is the aforementioned arc-like structure to the south-east, seen in emission from CH$_3$OH, $^{13}$CO and CH$_{3}$CN. Although weaker, a second plume like shape to the north-east of the continuum peak appears to be positioned as an almost inverse-mirror-image of the westerly plume, but it is only evident at slightly red-shifted velocities in the CH$_3$OH transition at 218.440\,GHz. Comparing to the CH$_{3}$CN K=3 emission, we find that there is a faint emission spot coincident with the strongest region of CH$_3$OH in the eastern plume (Fig.\,\ref{fig:fig5}, 23.2 to 23.9\,km\,s$^{-1}$, offset +1.5$''$,+0.4$''$) $-$ close to the OH maser spot.

The combination of the emission and velocity structure of these tracers and the OH maser emission all point to them tracing the cavity working surfaces where the wide-angle wind interacts with the inner edge of the dusty structure around G17.64.

\subsection{SiO and CH$_3$CN connection}
\label{siochconn}
\begin{figure}
\begin{center}
\includegraphics[width=0.52\textwidth]{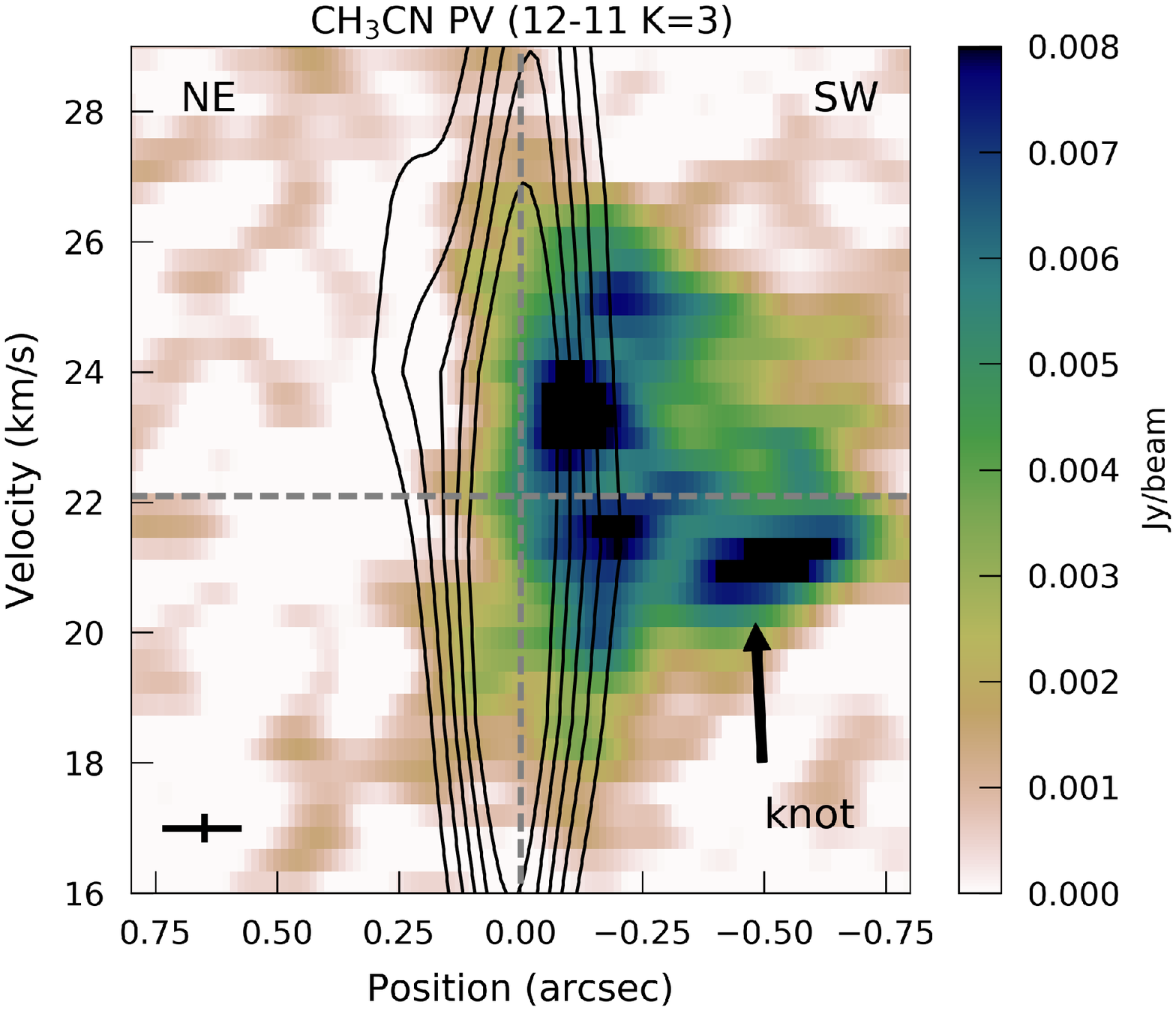}
\caption{Position-velocity (PV) diagram of the CH$_3$CN (J=12-11) K=3 line (colour map) overlaid with the PV diagram of the SiO as in Fig.\,\ref{fig:fig4} (black contours). There is notable emission to negative spatial offsets related to a south-west direction along the PV cut, whereas to positives offsets (north-east) there is barely any emission. Both blue- and red-shifted emission coincide to the south-west, related to the signature of either a wide-angle wind, with flows nearly radially, or that there is simultaneous infall from material in front and behind the edge of the disc. The emission at bluer velocities and high negative offset (-0.5$''$) is associated with the knot structure identified in Fig.\,\ref{fig:fig5}. The spatial and velocity resolution of the CH$_3$CN map is shown at the bottom left, while the direction along the PV cut to the NE and SW is indicated at the top.}
\label{fig:fig7} 
\end{center}
\end{figure}

In Fig.\,\ref{fig:fig7}, we show the PV diagram of the CH$_3$CN (J=12-11) K=3 emission extracted along the same cut as that for the SiO (NE to SW, see the disc major axis in Fig.\,\ref{fig:fig6}), however we use a width of 12 beams in order to encompass emission from the plume and the knot. The majority of the emission is at negative offsets, to the south-west, with only slight blue-shifted emission to the north-east ($+$0.2 offset). From an offset of -0.1 to  -0.3$''$ we see co-spatial blue- and red-shifted emission, as we might expect if there were a wide opening angle cavity where material flows almost along the line-of-sight closest to the source. Conversely, it could also be indicative of simultaneous infall at both the front and rear of the disc. However, infall would generally be understood to speed up as it comes closer to the star, whereas the brightest and strongest emission features from the plume indicate the opposite, the velocities (both blue- and red-shifted) increase from an offset of about -0.08$''$ to an offset of -0.28$''$. Thus the wind scenario, where material is continually driven away from the star in the plane of the disc, could generate such a PV diagram. Note, the PV diagram does indicate an overall triangle shape considering all the emission and shows generally higher-velocities closer to G17.64, and could be interpreted as an infalling stream. However, this assumes that all of the emission is from a spatially coherent structure, whereas in the channel map for CH$_3$CN (Fig.\,\ref{fig:fig5}) there is a physical break between the plume and knot. The knot emission spatially offset at -0.5$''$ is seen at bluer velocities and is disjoined from the main emission. Rather than being part of an infalling flow or stream, that contradicts with the plume only emission which increases in velocity with distance from G17.64, the knot is potentially emission from another structure, or given its location, due to an over-density where the wind impacts the proposed dust structure around G17.64. All the CH$_3$CN emission is within the low velocity spur regions at larger positional offsets when compared with the SiO emission. The CH$_3$CN could be indicative of the interface between a wide-angle wind blown from the putative disc with the surrounding diffuse medium.

\subsection{H30$\alpha$}  
\label{H30}
\begin{figure}
\begin{center}
  \includegraphics[width=0.52\textwidth]{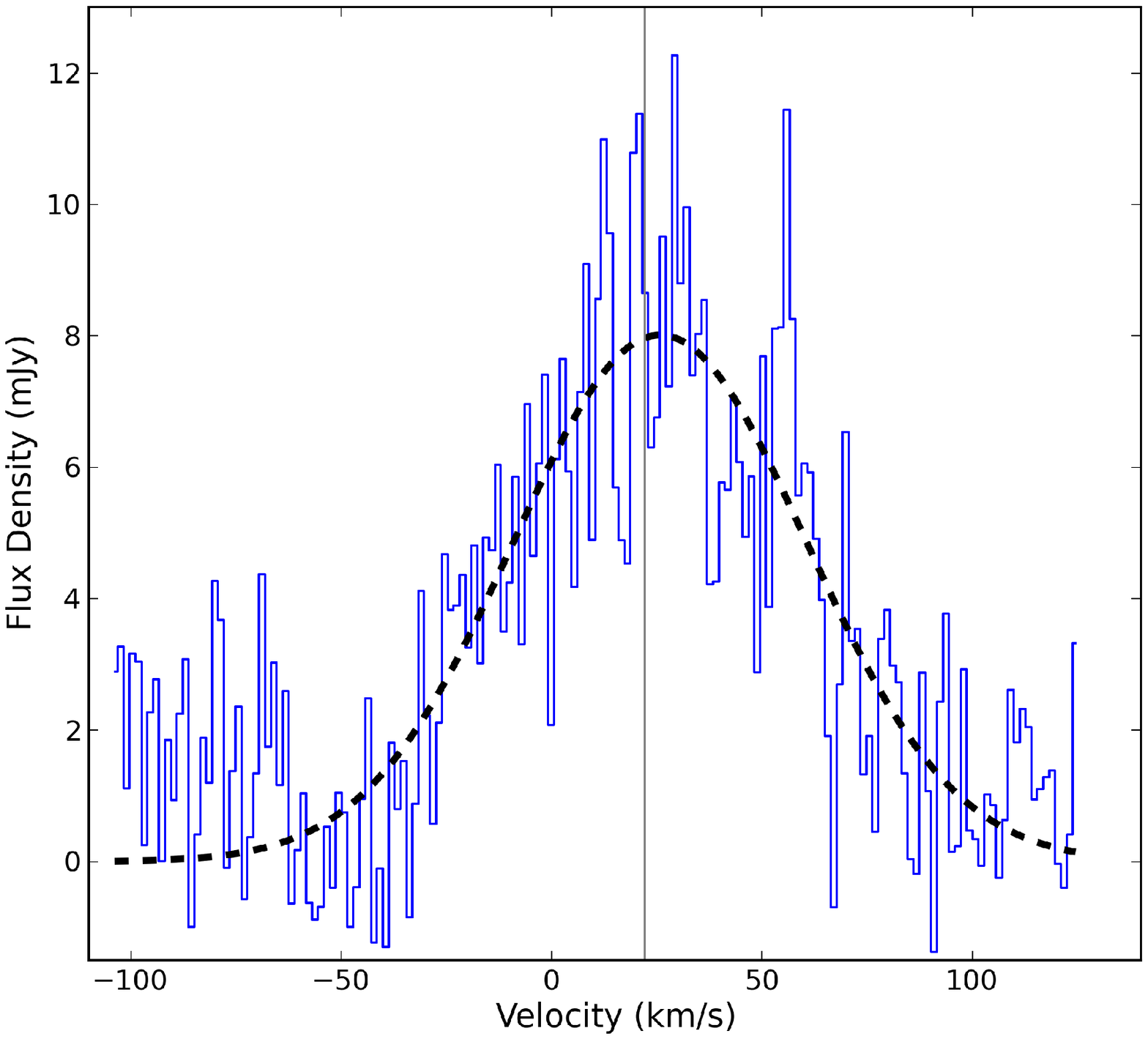}
\caption{Spectrum of H30$\alpha$ line extracted from within a single beam centred upon the continuum peak location over-plotted with a Gaussian fit (dashed black). The velocity resolution is 1.4\,km\,s$^{-1}$. The line is centred at 22.1\,km\,s$^{-1}$ the V$_{\rm LSR}$ of G17.64 as indicated by the vertical grey line. The base flux level for the Gaussian fit was fixed to zero. The line is weak compared to the other O-star sources \citep{Cesaroni2017} but is exceptionally broad, with a FWHM of $\sim$81.9$\pm$1.7\,km\,s$^{-1}$.}
\label{fig:fig8} 
\end{center}
\end{figure}

H30$\alpha$ hydrogen recombination line emission (231.9\,GHz) is detected from G17.64. It appears weaker than that from G29.96, G24.78 and G345.49, the other O-stars in our sample where it is also detected \citep{Cesaroni2017}. The emission from G17.64 is exceptionally broad ($\sigma$\,$\sim$33.5\,km\,s$^{-1}$, or FWHM$\sim$81.9$\pm$1.7\,km\,s$^{-1}$ from a Gaussian fit - Fig.\,\ref{fig:fig8}) and spatially unresolved. The baseline of the Gaussian fit was fixed at the zero flux level. G17.64 can be regarded as a broad-radio-recombination-line source \citep[c.f.][]{Sewilo2008, Kim2017}. We note that by eye there appears to be a narrow peak of emission at around 50\,km\,s$^{-1}$ which could tentatively be identified as HNCO v=0 28(1,2)-29(0,29), E$_{u} \sim$470\,K. Given the narrow width we do not consider it to significantly effect the broad line-width found for the H30$\alpha$ line.

Following the analysis outlined in \citet[][with electron and ion densities of 10$^7$\,cm$^{-3}$ and a temperature of 9000\,K]{GalvanMadrid2012} pressure broadening can at most explain $\sim$0.6\,km\,s$^{-1}$, while thermal broadening contributes $\sim$20\,km\,s$^{-1}$. Thus, if the line width we measure has only a typical small non-thermal turbulent contribution ($\sim$5\,km\,s$^{-1}$, \citealt{Sewilo2008}), there must be a significant underlying bulk gas motion component \citep[e.g.][]{Keto2008}, potentially comprised of infall (accretion), outflowing material (wind or expansion, c.f. \citealt{Moscadelli2018}) and/or rotation. We cannot rule out a combination of all mechanisms contributing to the bulk motion when we consider the emission is spatially unresolved in the current data. The high velocities and signpost of ionisation and high temperatures ties with the release of silicon and sulphur from the grains close to G17.64 due to associated shocks. The H30$\alpha$ emission will be addressed further in Klaassen et al. (2018 in prep.). 

\section{Analysis and Discussion}
\label{holder1}
\subsection{Modelling the kinematics of the compact disc emission}  
\label{mods}

We created a simple parametric model to further investigate the kinematics of the potential disc seen in SiO emission. This follows in the spirit of the models outlined in \citet{Richer1991} where a synthetic PV diagram is produced for comparison with the observations \citep[see also][]{Wang2012,Girart2017}. From the outset, given the limitations of the data in terms of spatial resolution and due to degeneracies with various parameters we do not aim to specify and fit every possible parameter, but rather address the qualitative physical structures and kinematics that are consistent with the data and their interpretation. We make a qualitative by-eye assessment for suitable matches with the data. The main parameter we aim to characterise is the stellar mass, as this is key to understanding whether G17.64 is truly an O-type YSO.

We build our model in a 3D environment where each volume element holds an emissivity ($\varepsilon(x,y,z)$) and a velocity ($V(x,y,z)$). We only assume an optically thin representation, and consider only two geometric structures in the model, a thin disc and a thin disc surface layer. The emissivity of the structures are parameterised by power-laws representative of the combination of density and temperature with radius, $\varepsilon \propto (r/r_0)^{p+q} $, where $r$ is the cylindrical radius, $r_0$ is an arbitrary normalisation radius, and $p+q$ is the `combined' power-law index \citep[e.g.][]{Brinch2010,Fedele2016}, where $p$ is the density power-law index and $q$ is the temperature power-law index. 
The disc, or disc surface can be independently parameterised by its inner and outer radii. The relative strength, or weighting, of the emission from each structure is also a free parameter, $\varepsilon_{\rm wei}$. The models are limited to two velocity field descriptions: (1) Keplerian rotation, $V_{\rm rot} = (GM/r_d)^{1/2}$, where $V_{\rm rot}$ is the velocity as a function of radius ($r_d$) in the disc plane, $G$ is the gravitational constant and $M$ is the stellar mass; and (2) radial motion in the disc plane. The magnitude of the radial motion is parameterised by the weighted free-fall velocity, $V_{rad} = A_{\rm wei}\,(2\,GM/r_d)^{1/2}$, where $A_{\rm wei}$ varies from 0.5 to 1, where 1.0 corresponds to pure free-fall. 

Our models are built from different combinations of structures and kinematics, as detailed in the sub-sections below. The principle is that we begin with the simplest possible description before including more complex scenarios with more free parameters.

For each volume element we establish the intensity of the emission by weighting with the emissivity profile, while the velocities are broadened by a Gaussian filter with a typical turbulent line width of 0.5\,km\,s$^{-1}$ at disc scales \citep{Richer1991}, although later we account for our observation spectral resolution. The model is inclined as required (an edge-on disc would be at 90$^{\circ}$) and the line of sight axis is integrated to make a synthetic datacube ($RA.,DEC.,VEL.$). The cube is finally convolved with a Gaussian kernel accounting for the spatial and velocity resolution before we extract the PV diagram. For comparisons with the observations, we scale the peak emission of the model PV diagrams to that of the data. Table \ref{table3} indicates the various model parameters and the possible ranges investigated. We explore a large parameter space with a relatively coarse model grid, again due to the limited resolution of the data at hand. Figure \ref{fig:fig9} shows the representative `best' matches to the kinematics of the observations from each model regime. The left plots indicate the PV plots as a schematic while the right plots show the models compared with the observations, top to bottom follows the models as presented in the following sub-sections.

\begin{table}[h!]
\begin{center}
  \caption{Parameters for the various models tested, the parameter ranges and the increments used.}
  {\footnotesize
\begin{tabular}{@{}lrrr@{}}
\hline
\hline
 Parameter &  &  Range   & Increment \\
\hline
\multicolumn{4}{c}{{\bf Disc - rotation model}} \\    
\hline
Stellar Mass  &  $M$            &  5$-$50    & 10$^a$           \\
Disc inner radius &  $R_{\rm inner}$            &  10$-$50  & 10             \\
Disc outer radius &  $R_{\rm outer}$            &  100$-$600  & 100               \\
Emissivity powerlaw &  $p+q$            &  -2.6$-$-1.8  & 0.2               \\
System Inclination &  $i$            &  30$-$90$^{\circ}$    & 20$^{\circ}$           \\

\hline
\multicolumn{4}{c}{{\bf Disc - rotation and radial model}} \\
\hline
Stellar Mass  &  $M$            &  10$-$30    & 10           \\
Disc inner radius &  $R_{\rm inner}$            &  20$-$50  & 10             \\
Disc outer radius &  $R_{\rm outer}$            &  100$-$600  & 100               \\
Emissivity powerlaw &  $p+q$           &  -2.6$-$-1.8  & 0.2               \\
Radial vel. weighting  & $A_{\rm wei}$   & 0.5$-$1.0 & 0.25  \\
System Inclination &  $i$            &  70$^{\circ}$    & fixed            \\

\hline
\multicolumn{4}{c}{{\bf Disc - rotation and disc surface radial model}} \\
\hline
Stellar Mass (M$_{\odot}$)  &  $M$            &  10$-$30    & 10           \\
Disc inner radius (au) &  $R_{\rm inner}$            &  20$-$50  & 10             \\
Disc outer radius (au) &  $R_{\rm outer}$            &  100$-$600  & 100               \\
Disc Surf. inner radius (au) &  $Rs_{\rm inner}$            &  50$-$600  & 100$^b$             \\
Disc Surf. outer radius (au) &  $Rs_{\rm outer}$            &  650  & fixed               \\
Emissivity powerlaw &  $p+q$           &  -2.6$-$-1.8  & 0.2               \\
Emissivity weighting & $\varepsilon_{\rm wei}$  &   0.5$-$2.0  &  0.5 \\
Radial vel. weighting  & $A_{\rm wei}$   & 0.5$-$1.0 & 0.25  \\
System Inclination ($^{\circ}$)&  $i$            &  70$^{\circ}$    & fixed            \\

\hline
\label{table3}

\end{tabular}
}
\end{center}

  {\footnotesize
     $^{\rm a}$ Between 5 and 10\,M$_\odot$ the increment is 5\,M$_{\odot}$ and 10\,M$_{\odot}$ thereafter\\
  $^{\rm b}$ Between 50 and 100\,au the increment is 50\,au, and 100\,au thereafter.}

\end{table}

\subsubsection{Disc only - rotation}
Using models where rotation is the only velocity component we are able to match only two opposite quadrants in PV space, exactly as in Sect.\,\ref{Ssio}, even when accounting for the resolution of the observations (Fig.\,\ref{fig:fig9}, top panel). We can match the highest velocities in the PV diagram using a small radius disc, but this model does not account for the low velocity spurs. A large outer radius disc with a respectively larger inner radius (vs. a small disc) rotating in the opposite sense does match the spurs, but cannot describe the high velocity emission. A considerable number of disc-only models represent the data (in two of the four quadrants) equally well (or equally poorly) primarily due to the degeneracy between stellar mass and inclination angle. Given the modelling undertaken previously by \citet{dewit2011} we choose to fix the inclination angle to 70$^{\circ}$ (where 90$^{\circ}$ is edge-on to the disc). A side effect is that the stellar mass is a lower limit in all models considering the close to edge-on orientation of the disc.  

With a fixed inclination the aforementioned high velocities are represented by discs with $R_{\rm inner}$ = 20-50\,au and $R_{\rm outer}$ = 100-600\,au for stellar masses ranging from 10$-$30\,M$_{\odot}$ in various parameter combinations. For larger masses the velocities at all radii become too great, and although we could increase the inner radius beyond 50\,au the distribution of emission shifts away from low spatial offsets which is inconsistent with the data. The lowest mass of 5\,M$_{\odot}$ cannot provide sufficiently high velocities even with a 10\,au inner radius (and such a low mass source is anyway inconsistent with the luminosity of G17.64). We therefore rule out M$<$5\,M$_{\odot}$ and M$>$30\,M$_{\odot}$ stellar masses. There is however still a degeneracy between mass and $R_{\rm inner}$. Convolution of the model with the observable spatial and velocity resolution acts to reduce the high velocity emission peaks by smoothing the emission and also results in higher-mass larger-inner radius models appearing the same as lower-mass lower-inner radius models. A disc rotating in the opposite direction, matching the spurs, is well represented by a more limited range where $M$ = 10\,M$_{\odot}$, $R_{\rm inner}$ = 20\,au, and $R_{\rm outer}$ = 400$-$600\,au. A larger stellar mass in this case results in too high velocities at large spatial offsets. Variation of the emissivity power law has a marginal effect on the distribution of the intensity in the PV diagram. Steeper profiles (-2.6 to -2.2) produce more emission from smaller radii that have the largest velocities and thus begins to create two emission peaks, inconsistent with the observed intensity distribution. None of these models match the entire PV plot. 

\subsubsection{Disc only - rotation with radial motion}
\begin{figure*}
\begin{center}
\includegraphics[width=0.85\textwidth]{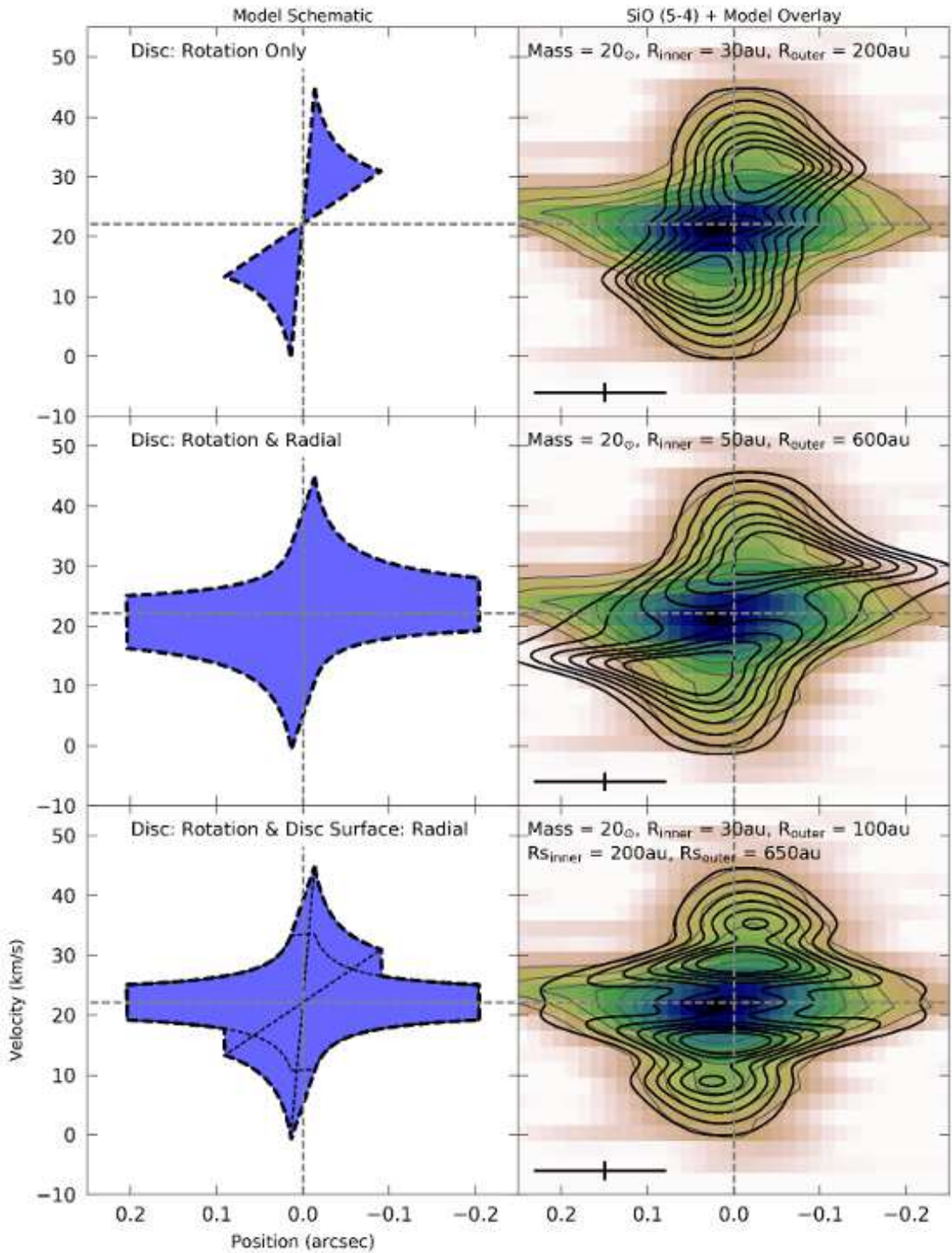}  
\caption{Plots indicating the various models tested as a schematic (left) and then as an overlay on the observed SiO data (right). Top to bottom shows: the disc model with only rotational velocities; the disc model with rotation and radial velocity motion; and the disc with disc surface model where the disc has rotational motion whereas the surface has radial motion. Both disc only models (top and middle) have intensity profiles where most emission is shifted to high velocities and small spatial offsets, unlike the centrally peaked observations. For the disc model with rotation and radial motion (middle), the observer sees the vector sum of the velocity component, such that there is always a resulting velocity offset at larger spatial offsets due to the rotational component. This does not match the data. For models of a rotating disc combined with a disc surface that has radial motion (bottom), the velocities from each structure are superposed and thus can represent the entire observed PV diagram (the bottom-left plot shows the Keplerian and radial velocities profiles in the dotted thin line). In all right hand plots the data resolution is shown to the bottom-left while the contours for the data and model are from 20 to 80\,\% in 10\,\% steps of the peak emission. Note the basic models parameters shown in the right plots.}
\label{fig:fig9} 
\end{center}
\end{figure*}

In an attempt to reproduce the emission in all four quadrants, we now include radial motion, interpreted as accretion through, or expansion of the disc, along with rotation. Each volume element in the model now has an assigned radial and rotational velocity component (Fig.\,\ref{fig:fig9}, middle panels). Following from the above models we limit the mass range between 10 and 30\,M$_{\odot}$. We find a wide range of parameters that somewhat match the observed PV diagram partially, however, the disc is required to have an outer radius ($R_{\rm outer}$) of at least 600\,au in order to fit the spurs. The addition of the radial motion actually narrows the model parameter space, requiring an inner radius between 30$-$50\,au to better reproduce the observables, although it is still degenerate with stellar mass. The weighting of the radial to rotating velocity components provides an additional degeneracy, although the parameter range of the `best' matching models is already broad.

With a fixed outer radius $R_{\rm outer}$ of 600\,au, the 10\,M$_{\odot}$ stellar mass models provide a reasonable match for $R_{\rm inner}$ = 30-50\,au, and various combinations of disc power-law profile and radial to rotational velocity weightings. For a 20 \,M$_{\odot}$ stellar mass the parameters are further restricted, $R_{\rm inner}$ = 40-50\,au, power-law profile $>$-2.2, and A$_{\rm wei}$ = 0.5$-$0.75, while for a 30 \,M$_{\odot}$ central mass there is only one parameter set that matches the observed PV, $R_{\rm inner}$ = 50\,au, power-law = -1.8 and A$_{\rm wei}$ = 0.5. 

Although these models have a more restrictive range, owing to the inclusion of radial motion compared to the rotation only models, the major problems are the considerable rotational velocities up to 5$-$6\,km\,s$^{-1}$ at the largest spatial offsets in the spurs ($\pm$0.23$''$, i.e. $>$500\,au radii) and the intensity distribution at low spatial offsets being biased to larger velocities than indicated in the observations. What we do deduce is that the high velocity components matching the skew in the PV diagram can only be provided by a disc rotating in one direction (e.g. Fig.\,\ref{fig:fig4}, left and centre), as a large disc rotating in the opposite sense (previously matching only the spur emission - e.g. Fig.\,\ref{fig:fig4}, right) is ruled out as the inner radius is too large in these cases and always under-represents the high velocities while still not fully matching the spurs. Again, none of the models are entirely consistent with the observed PV diagram.

\subsubsection{Disc and Disc Surface}
We now consider that the large ($>$400\,au) spatial offset emission close to V$_{\rm LSR}$ velocities must be provided by a structure separated from the rotating disc. We model this by including a thin disc surface with only radial motion with a variable weighting, as outlined above (Fig.\,\ref{fig:fig9}, bottom). This motion can be interpreted as either accretion onto the disc or a wind blowing away the disc. More realistically, there is likely a stratification of the disc in scale height where the velocity transitions from rotation to radial motion, although constraining this detail is beyond the scope of our modelling. Here we fix the scale height of the surface layer to only one resolution element above and below the disc (10\,au in our models) and note that is spatially separate from the thin-disc (in the mid-plane). This thin surface layer is the limiting case as we cannot constrain scale height information with the data at hand. Because the structures are separate, the corresponding kinematics are superposed in the final model PV diagram (Fig.\,\ref{fig:fig9}, bottom-left). The emissivity profile of the disc surface is exactly the same as the disc, although it can be weighted by a factor between 0.5 and 2.0, $\varepsilon_{\rm wei}$.

There are obvious degeneracies in these models given the additional flexibilities of the independent disc inner and outer radii and disc surface inner radius along with the weightings of intensity and velocity components. We fix the disc surface outer radius to 650\,au to represent the large offset emission near V$_{\rm LSR}$ velocities. Again, a range of models match the observations. For a 10\,M$_{\odot}$ stellar mass the models are limited to only $R_{\rm inner}$ = 20\,au as they otherwise do not reproduce the high-velocity emission, while $R_{\rm outer}$ < 200\,au so as not to spread too much emission to the lower-velocity regions at larger radii (i.e. to the top-right and bottom-left quadrants). The disc surface inner radius, $Rs_{\rm inner}$, has to be $<$ 200\,au in general so as to provide emission from radial motion in the top-left and bottom-right quadrants of the PV plot. Increasing the stellar mass to 20\,M$_{\odot}$ we find the same degeneracies as before, the inner radius of these models is $R_{\rm inner}$ = 30$-$40\,au while $R_{\rm outer}$ $<$200\,au and $Rs_{\rm inner}$ $<$200\,au, as for the 10\,M$_{\odot}$ case. At the final mass tested, 30\,M$_{\odot}$, again $R_{\rm inner}$ must increase to 40$-$50\,au, $R_{\rm outer}$ must remain $<$200\,au while $Rs_{\rm inner}$ is between 100$-$400\,au in general, but not well constrained. Notably, for this highest stellar mass the match with the data appears slightly worse, primarily as the slope between high-velocity small-offset emission is slightly steeper compared with the lower-mass models. For all the models, in general the other parameters can span essentially the full ranges, although importantly all models provide a qualitative match to the entire observed PV plot in all quadrents, depending on the parameter combination.

Overall the 10-20\,M$_{\odot}$ stellar mass models best represent the data, although a number of 30\,M$_{\odot}$ stellar mass models are still reasonable. We do rule out masses between 40 and 50\,M$_{\odot}$ as they have too large velocities at all spatial offsets, as discussed in the disc only models, although we cannot further constrain the stellar mass. Comparing with the disc only models, the disc and disc surface models do provide a qualitative match to the entire PV diagram, and moreover, indicate that two velocity components from different structures are required. With higher resolution observations these structures could possibly be disentangled. Taking into consideration the luminosity of G17.64 (1$\times$10$^5$\,L$_{\odot}$) and that it appears as the main dominant and isolated source in continuum emission, a mass in the upper limit of the matching model range $\sim$20\,M$_{\odot}$ is more probable.

\subsection{Possible origins of the SiO emission and kinematics}
\label{origns}
The presence of SiO means silicon must have been released into the gas phase by sputtering, likely due to a C-shock \citep[e.g.][]{Schilke1997, Gusdorf2008} or by grain-grain collisions, shattering or vapourising the grains \citep{Jones1996,Guillet2011}. The maximal velocities of the SiO we detect (relative to the V$_{\rm LSR}$) are $\sim$20\,km\,s$^{-1}$ marginally slower than the typical C-shock speeds $>$25\,km\,s$^{-1}$ \citep{Caselli1997}. Observationally our SiO detection is only resolved by approximately three to four beams and it is plausible that the even higher velocity emission is beam diluted in the spatially unresolved central region, and these are also the projected velocities to our viewing angle. The spatially unresolved H30$\alpha$ certainly has a broad enough velocity component, indicative of a fast expansion, which would be sufficient for the release of silicon in C-shocks. However, low velocity shocks are also consistent with SiO in other high-mass regions \citep[e.g. 7-12\,km\,s$^{-1}$][]{Louvet2016}, and could also provide the means of releasing SiO into the gas phase. In our model scenarios a combination of rotation and radial motion are required, but radial motion of either infall or expansion cannot be distinguished as both provide the same PV pattern in the optically thin case. We argue that an expansion scenario is more plausible for the kinematics, as explained in the following.

We present three qualitative scenarios for the possible origin of the SiO and tie these with the kinematics.
\begin{enumerate}
\item In the case of an infalling accretion shock to the outer edge of the disc at $\sim$600\,au radii from accretion streams, we have to speculate that the plume structures seen in CH$_3$CN and CH$_3$OH are accretion flows feeding the disc (ignoring the outflow and wind intepretation of the $^{13}$CO and CH$_3$OH and the maser emission, as previously discussed). The velocities seen in SiO are low ($\pm$2-4\,km\,s$^{-1}$ from the V$_{\rm LSR}$) where the CH$_3$CN plume meets the disc outer radius (see Fig.\,\ref{fig:fig7}). At the location of the strongest south-westerly plume emission the velocities are marginally shifted by only $\pm$2-3\,km\,s$^{-1}$. Thus very low velocity accretion `shocks' would be the inferred release mechanism for the silicon, leading to SiO production at outer disc radii beyond $\sim$600\,au. These shock speeds are lower than those suggested in other low velocity shock regions \citep{Louvet2016} and inconsistent with C-shocks. In this scenario, if SiO were created, it would be accreted through the disc from the outer edge and thus provide the rotation and radial infall kinematics. 
  
\item Observationally, the H30$\alpha$ detection points to a hot, compact, ionised region with a very broad velocity range. The plume emission can easily be understood as a wide-angle wind - and the knot where the wind impacts the surrounding medium (in CH$_3$CN). We propose then that the silicon is released in fast shocks ($>$25\,km\,s$^{-1}$) at the inner edge of the disc where we have evidence for such high velocities, and potentially at the disc surface interface due to an expansive wind driven by G17.64. Here, the SiO produced is in rotation at small radii while the disc surface is free to expand away in a disc wind. This scenario produces the kinematics observed, and is preferred by our modelling.
 
\item Silicon could be released within a rotating disc structure by shocks at a centrifugal barrier close to the star \citep[e.g. 50$-$100\,au, ][]{Sakai2014,Oya2016} within the disc. Indeed we could represent the high velocity linear structure in the PV diagram by an unresolved SiO ring within a larger disc \citep[c.f. the SO ring observed by ][]{Sakai2014}. However, to explain the low velocity SiO emission at large disc radii is difficult in this third scenario. This is because the SiO is produced at small radii ($<$100$-$200\,au), at the centrifugal barrier in this scenario, and therefore cannot simultaneously be infalling onto the disc from larger radii (600$-$700\,au). The lower velocity larger scale emission is only easily explained by material being blown to larger radii from the disc surface by a wind. 
\end{enumerate}

In addition, we also consider previous radio-wave data. \citet{Menten2004} state that accretion shock models \citep{Neufeld1996} cannot replicate the radio fluxes found for G17.64 (i.e. our scenario 1). On the other hand the spectral index ($\alpha$=1.2) of the emission $<$43\,GHz points to a compact stellar wind that must be an accelerating flow or have recombination in an expanding flow \citep{Reynolds1986}. Our presented observations confirm the recombination requirement. Furthermore, the fluxes reported by \citet{Menten2004} at 8.4\,GHz are entirely consistent with a wind nature \citep[see][]{Anglada2015,Purser2016}. Although with our data we cannot rule out the case of infall towards G17.64, especially on small spatial scales $<$50\,au based on evidence from CO absorption lines \citep{Mitchell1990} and H$_2$O red-shifted masers \citep{Menten2004}, we can rule out infall as producing SiO in our scenario (1). The small scale accretion markers cannot explain infall at $\sim$650\,au radii where shocks are required and where we detect SiO. Furthermore, we believe that it is highly improbable that the plumes are accretion flows due to the association with outflow tracers and because the CH$_3$CN velocity profile from the strongest, coherence plume emission does not show an increase at lower offsets as might be expected from infall (Fig.\,\ref{fig:fig7}). Cases (2) or (3) are the preferred scenarios for the production of SiO, and although our data cannot provide constraints on which method is most prevalent, we can state that in either scenario we require an expanding velocity component in addition to rotation, both agree with the observables and both are also consistent with literature data.

\subsection{G17.64 in context}  
\label{context}

\begin{figure}
\begin{center}
\includegraphics[width=0.48\textwidth]{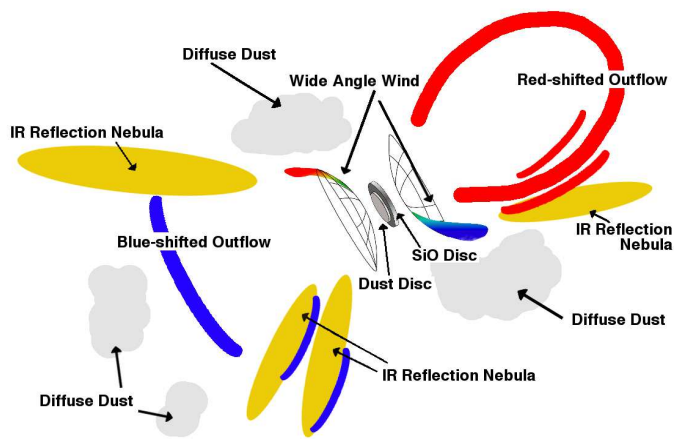}  
\caption{Schematic view of G17.64 and the surrounding region (not to scale). All main features are identified with labels. The dust disc is drawn to have a smaller size compared with the SiO disc in line with the presented observations as we resolve the SiO emission but not the dust emission. For the wide-angle wind the colours indicate the rough velocity of the molecular tracers we detect at the interaction with the inner edge of the dust structures. The emission to the west is blue shifted, tracing the wide-angle wind directed almost directly towards us, while in the east it is red-shifted as we see emission from the back edge of the cavity. In addition, note the arc shape of the blue-shifted emission to the south-east which is traced by $^{13}$CO (see Fig.\,\ref{fig:fig2}), CH$_3$CN and CH$_3$OH. For orientation north is up and east is left.}
\label{fig:fig10} 
\end{center}
\end{figure}

The kinematic signatures of the SiO emission coupled with our modelling supports that there is a compact ($<$200\,au radii) rotating disc structure around at least a 10\,M$_{\odot}$ central source which is driving a wide-angle wind in the disc plane that then redirects into a large-scale outflow bubble. We confirm the previously predicted disc$-$outflow structure of the region based on earlier IR observations at larger scales \citep{Kastner1994,Murakawa2008,Minchin1991}. In Fig.\,\ref{fig:fig10} we present our schematic view of G17.64, indicating all features discussed in this work. It shows the CO outflow as the red-bubble and the blue arc structure and the relation with the IR reflection nebula, probably tracing the cavity walls \citep[see also Fig.\,2 in][]{Murakawa2008}, while we also add the compact disc structure and the wide-angle wind.  

\subsubsection{Comparing with other O-type YSOs}
Crucially, matching the high spatial resolutions of our data probing $<$500\,au, we compare G17.64 with: the other two (of five) close ($<$3\,kpc, resolution 0.2$''$, i.e. $<$600\,au) regions harbouring O-type YSOs observed in our original sample \citep{Cesaroni2017}; the proto-O-type YSO W33A MM1 (d $\sim$2.4\,kpc, resolution 0.2$''-$0.3$''$) observed by \citet{Maud2017}; the clustered NGC\,6334\,I system from \citet[][d $\sim$ 1.3\,kpc, resolution $<$0.2$''$]{Hunter2017}; the long baseline observations of O-stars in W51e2 and W51n (\citealt{Goddi2018}, d$\sim$5.4\,kpc, resolution 0.03$''-$0.04$''$); and G328.2551-0.5321 (hearafter G328) observed by \citet{Csengeri2018} at 400\,au resolution.

On most accounts, G17.64 stands out clearly as being isolated, in a region of low fragmentation of strong cores and being highly compact in terms of the dust continuum emission and SiO. Typically, the other sources drive more linear or `V' shaped outflow or jet structures in SiO, while their continuum emission shows at least three to four other compact cores within 5000\,au of the main massive YSO (except G328 which also appears to be isolated). Moreover, some of the other sources have evidence of filament-like streamers, for example W33A, W51e2 and W51n, suggestive of accretion flows and generally have more complex molecular species detected over larger regions ($>$5000-10000\,au). The latter is a signpost of more dense material which has been warmed relatively recently or at least not exposed to harsh ultra-violet or ionising radiation which destroys complex molecules. 

\subsubsection{Why is there a detectable disc in G17.64+0.16 - Isolation or evolution?}
Taking the single case of G17.64, it may only have an unperturbed and observable disc due to its isolated nature, forming in an environment dominated by its own gravitational potential and away from other energetic events (e.g. away from other massive YSOs forming nearby, H{\sc ii} regions or jets and outflows). Indeed the data we present portrays G17.64 as a single, dominant source within the ALMA primary beam ($>$50000\,au in diameter, 0.24\,pc), but also in the IR it is the single brightest source (excluding the nebula emission) within at least a 25000\,au (0.12\,pc) diameter region. Extrapolating from this idea, we see that G328 also appears isolated at mm wavelengths \citep{Csengeri2018} and is one of the aforementioned comparison O-stars that has strong indications of an accretion disc, although it also drives a very collimated outflow, suggesting it is rather young. The other comparions sources in much more complex regions, and do not show disc evidence.

At lower spatial resolutions ($>$1000\,au) other regions with complex dust continuum and molecular line emission do show large-scale rotation consistent with disc-like structures (e.g. AFGL 4176 $-$ \citealt{Johnston2015}, G11.92-0.61 $-$ \citealt{Ilee2016,Cyganowski2017}, W3(H$_2$O) $-$ Ahmadi et al. 2018 accepted.), as does G31.41$+$0.31, a more distant source part of our O-star sample \citep{Beltran2018}. These targets may still further fragment if observed at higher resolution. This appears to be the case at 3\,mm wavelengths for G31.41+0.31 (Beltr\'{a}n private communication) and G11.92-0.61 (Ilee private communication) which have recently been observed with ALMA using long baselines ($>$10\,km). It is unclear whether the discs are dominated by a single source, or a common circumbinary disc. Indeed fragmentation of a thought-to-be single source was recently seen in the massive YSO source NGC7538 IRS1, observed at $\sim$150\,au resolution by \citet{Beuther2017}. These authors now distinguish two HCH{\sc ii} regions thought to be surrounded by individual disc structures but both encompassed by a larger scale circumbinary rotating structure \citep[see also][]{Goddi2015}. \citet{Maud2017} found that their target, W33A MM1, fragments into at least 6 continuum sources at sub-500\,au scales whereas previous, lower-resolution, SMA data pointed initially to a single massive YSO \citep{GalvanMadrid2010}, while \citet{Beuther2017} find at least two, strong continuum sources in G351.77-0.54 within $\sim$500\,au (ALMA 0.43\,mm observations with 0.06$''$ resolution - 120\,au) but find no clear Keplerian disc signatures. G17.64 is uniquely isolated from other strong compact cores and lacks continuous extended emission, although the possibility of an unresolved close $<$400\,au binary, or fragmentation below these scales cannot be ruled out with the current ALMA data. 

  Detection of the disc around G17.64 could also be aided by an evolutionary progression. G17.64 exhibits bright IR emission, which suggests that at least part of the natal cloud has been cleared enough to lower the opacity at short wavelengths, rather than the IR emission being due to a favourable viewing down the outflow cavity, as in IRAS\,13481-6127 \citep{Kraus2010}. There is no longer a very dense obscuring envelope or opaque large-scale dust and molecular gas structures enclosing G17.64 as there is in the aforementioned comparison sources. The outflow from G17.64 is also a well developed bubble \citep{Beuther2005,Vaidya2011} and not a linear jet \citep[e.g.][]{Curiel2006}, while there is also spatially compact radio and broad line-width H30$\alpha$ emission that is expected only in the later stages of formation. In the context of a protostellar evolutionary path \citep{Hosokawa2009,Hosokawa2010,Kuiper2013}, the stellar photosphere of G17.64 would be contracting from a bloated phase, where the stellar surface was previously swollen by high-accretion, towards a main-sequence stellar configuration with a radiative outer envelope that is hot enough to begin the destruction and ionisation of the surrounding environment. This scenario points to G17.64 as being similar to S140 IRS1, which has a $\sim$100\,au dust disc, distinct lack of molecular line emission and is the prototype ionised disc wind source \citep{Hoare2006,Maud2013b}. Even thought there is evidence that some of the natal envelope has been dispersed very close to G17.64, it is plausible that G17.64 has finished its major accretion phase and that this may actually have allowed conditions to change such that a disc was able to stabilise. Revisiting G17.64 with the high sensitivities offered by the upgraded JVLA could shed light on an ionised disc wind nature while ALMA long baselines are required to resolve the continuum and SiO in the mm regime.

  For G17.64 it is also interesting that CH$_3$CN, a tracer commonly used to identify discs around massive YSOs, does not trace a disc. Certainly the sub-1000\,au resolution of these observations could provide a partial answer. When we compare with CH$_3$CN disc candidates presented in \citet{Beltran2016} we find that the spatial resolutions generally probe $>$1000\,au scales and thus are sensitive to the large-scale surrounding material. AFGL 4176 observed by \citet{Johnston2015} is the clearest case of a coherent Keplerian-like disc traced in CH$_3$CN around a massive source, while \citet{Ilee2016} show evidence of rotation in CH$_3$CN and many other lines for G11.92. It remains unclear if CH$_3$CN would be the best tracer for smaller scale structures if they were observed with a more comparable resolution to our data, around four times better. \citet{Maud2017} show that CH$_3$CN in the complex region of W33A MM1 essentially traces all the hot ($>$100-200\,K) material and that it is not specifically associated with a disc when probing $<$1000\,au spatial scales. Previously, low resolution observations of W33A by \citet{GalvanMadrid2010} did point to a disc as they were not sensitive to the sub-structure. Indeed, in our overview paper of proto-O-stars we do not clearly find CH$_3$CN Keplerian discs in all sources \citep{Cesaroni2017}.

  Source evolution could also play a key role in not finding CH$_3$CN in a disc. If we consider G17.64 to be slightly more evolved, accounting for the previously mentioned reduction in complex chemical species and the H30$\alpha$ detection, then it is possible that molecules such as CH$_3$CN are now being destroyed close to G17.64. Other species, such as SiO, may then be preferentially detected in a disc that is no longer forming, but being ablated by a disc wind. Recent work by \citet{Girart2017} also suggests that the disc source, IRAS 18162-2048 (also known as GGD27 MM1 and responsible for the HH 80-81 jet) could also be at a later evolutionary stage. It also lacks bright complex organic molecular emission but is rich in sulphur bearing species, and has some HCN and CH$_3$OH emission, all consistent with being more evolved than a typical hot core and more similar to UCH{\sc ii} regions \citep{Minh2016}. IRAS 18162-2048 is also relatively isolated like G17.64, although it is a lower mass B0-type YSO.

  Finally, it is also worth discussing Orion Source I (Src I), which has previously been compared to G17.64 in the literature due to similar radio-wave properties, although G17.64 is an order of magnitude stronger given similar physical sizes \citep{Menten2004}. Intriguingly, Src I\footnote{There is some dispute whether Src I is a more evolved system, \citet[see][]{Baez2018}} is the only other known massive ($>$8\,M$_{\odot}$) YSO with SiO tracing a rotating disc structure \citep{Hirota2017}, but it also has SiO masers \citep{Goddi2009} and a larger scale rotating SiO outflow \citep[e.g.][]{Plambeck2009}. Long baseline ALMA observations were presented by \citet{Ginsburg2018}, who indicated that a number of molecular lines trace the rotating disc and disc surface structures. They refine the mass of Src I to $\sim$15\,M$_{\odot}$. In light of this, G17.64 also appears somewhat similar to Src I in terms of the unique SiO disc emission, however Src I would be positioned as a slightly younger source accounting for the considerably weaker radio emission, given a similar stellar mass, whereas G17.64 also has H30$\alpha$ radio recombination emission and is thought to be driving a radial wide-angle wind, not one that is co-rotating with the disc. 

\section{Conclusions} 
\label{conc}
We observed the luminous massive-YSO, G17.64+0.16 (AFGL 2136), at 1.3\,mm wavelengths with a resolution of $\la$0.2$''$ to probe scales below 400\,au. We find the source is isolated and essentially unresolved in strong dust continuum emission. The deconvolved fit parameters are consistent with the dust emission arising from a disc-like structure almost perpendicular to the large-scale molecular outflow traced here by $^{13}$CO. The $^{13}$CO emission, although partially resolved out, highlights the strongest emission regions that appear to trace the bubble-shaped walls of a cavity carved by the outflow itself and a wide-angle wind driven by G17.64. Emission from both CH$_3$CN and CH$_3$OH trace arc-shaped plumes of emission that curve away from the disc plane. The plumes coincide with maser emission and trace the inner working surfaces of the outflow cavity, confined by diffuse dust emission from a mostly dispersed or resolved out remant toroid or dark lane structure. We find strong SiO emission with an exceptionally broad velocity profile that is morphologically elongated almost perpendicular (PA$\sim$30$^{\circ}$) to the large-scale CO outflow (PA$\sim$135$^{\circ}$), inconsistent with a typical jet origin. Using parametric models to compare with the SiO emission we show that the kinematics are consistent with Keplerian rotation of a disc around an object between 10$-$30\,M$_{\odot}$ in mass, but crucially, the models must also have a radially expanding component from a separate structure. The radial motion component can be interpreted as a disc wind from the disc surface, in order to match the data. The upper mass range of the models and source luminosity are consistent with the picture of G17.64 as an O-type YSO. The detection of compact H30$\alpha$ emission further supports this picture as G17.64 is ionising the central, spatially unresolved, region and can drive the disc wind. The unique combination of the small radius disc $<$200\,au, the disc wind and the compact ionised emission all point to G17.64 being in the final stages of formation contracting from a bloated star to a main sequence configuration. This picture is in contrast to other relatively younger massive YSOs that have evidence of larger rotating structures, complex accretion streamers and rich molecular environments. The dispersal of surrounding material and the isolation of G17.64, and other comparison sources, along with a later stage of evolution, may be key requirements for discovering discs around similar massive YSOs.

\begin{acknowledgements}

  We thank the referee for their comments that helped to clarify a number of points within the paper.
  
  MSNK acknowledges the support from Funda\c{c}\~ao para a Ci\^{e}ncia e Tecnologia (FCT)
through Investigador FCT contracts IF/00956/2015/CP1273/CT0002, and the H2020 Marie-Curie
Intra-European Fellowship project GESTATE (661249).

  RGM acknowledges support from UNAM-PAPIIT programme IA102817. 

  ASM acknowledges support from the Deutsche Forschungsgemeinschaft (DFG) via the Sonderforschungsbereich SFB 956 (project A6).

  JCM, HB and AA acknowledge support from the European Research Council under the European Community’s Horizon 2020 framework programme (2014-2020) via the ERC Consolidator grant `From Cloud to Star Formation (CSF)’ (project number 648505).

  RK acknowledges financial support via the Emmy Noether Research Programme funded by the German Research Foundation (DFG) under grant no. KU 2849/3-1.

  V.M.R. has received funding from the European Union's H2020 research and innovation programme under the Marie Sk\l{}odowska-Curie grant agreement No 664931.

  This work was partly supported by the Italian Ministero dell\'\,Istruzione, Universit\`a e Ricerca through the grant Progetti Premiali 2012 -- iALMA (CUP C52I13000140001), and by the DFG cluster of excellence Origin and Structure of the Universe (http://www.universe-cluster.de, www.universe-cluster.de).
  
  This paper makes use of the following ALMA data: ADS/JAO.ALMA\#2013.1.00489.S and ADS/JAO.ALMA\#2016.1.00288.S. ALMA is a partnership of ESO (representing its member states), NSF (USA) and NINS (Japan), together with NRC (Canada), NSC and ASIAA (Taiwan), and KASI (Republic of Korea), in cooperation with the Republic of Chile. The Joint ALMA Observatory is operated by ESO, AUI/NRAO and NAOJ.

\end{acknowledgements}
\bibliographystyle{aa}

\appendix
\section{ACA $^{13}$CO outflow}
\label{AppendixA}
The CO outflow associated with G17.64 was previously observed and mapped by \citet{Kastner1994} in the CO (1$-$0) and (2$-$1) transitions. \citet{Maud2015} also targeted G17.64 in the $^{12}$CO, $^{13}$CO and C$^{18}$O (3$-$2) transitions as part of their survey of 99 massive-YSOs observed with the JCMT at $\sim$15$''$ angular resolution. From these single dish studies the outflow was seen to have strong blue-shifted emission to the south-east and a strong red-shifted lobe to the north-west. The blue-shifted lobe also aligns reasonably with the IR reflection nebula thought to highlight the cavity walls \citep{Kastner1992, Murakawa2008}. 
We use the analysis presented in \citet{Maud2015} adapted to the $^{13}$CO (2$-$1) transition observed with the ACA. We also use their estimates for the excitation temperature (T$_{ex}$ = 21.2\,K) and correct optical depth using a bulk average ($\tau_{13_{CO}}$ = 6.34). The resulting masses of the blue- and red-shifted lobes (integrated between the range indicated by \citet{Maud2015}, 6.0 to 20.0\,km\,s$^{-1}$ and 25.9 to 38.2\,km\,s$^{-1}$) corrected for optical depth are 1.64 and 8.40\,M$_{\odot}$ respectively. Having only relatively low sensitivity $^{13}$CO emission in the ACA observations, we only detect outflow line-wings to maximal velocities of $\sim$16.9\,km\,s$^{-1}$ to 35.6\,km\,s$^{-1}$ for the blue- and red-shifted lobes when comparing to \citet{Maud2015} who established their velocity ranges with $^{12}$CO.

\begin{figure*}[ht!]
\begin{center}
  \includegraphics[width=0.8\textwidth]{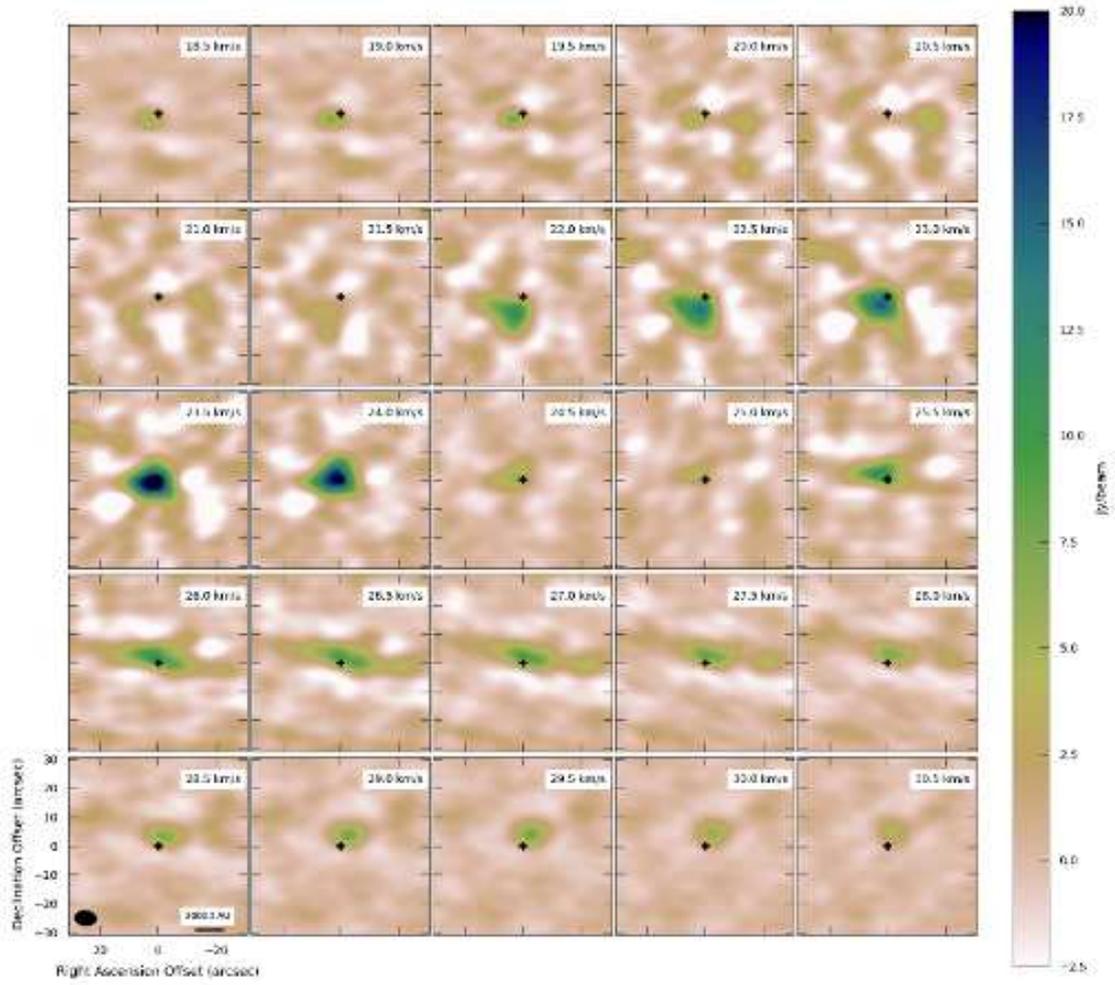}
\caption{Channel map of the ACA $^{13}$CO emission ranging from 18.5 to 30.5\,km\,s$^{-1}$ in steps of 0.5\,km\,s$^{-1}$. Recall the source V$_{\rm LSR}$ is 22.1\,km\,s$^{-1}$. Note the highest velocities of the outflow in the blue- and red-shifted emission occur to the south-east and north-west respectively, perpendicular the putative disc. The blue-shifted emission is detected down to 16.9\,km\,s$^{-1}$ but is co-spatial with that shown at 18.5\,km\,s$^{-1}$, while the red-shifted emission is detected to 35.6\,km\,s$^{-1}$ but is the same structure as the panel at 30.5\,km\,s$^{-1}$.}
\label{fig:fig11} 
\end{center}
\end{figure*}

We do not recover the masses as reported from the aforementioned single dish study (56.6 and 40.7\,M$_{\odot}$ for the blue- and red-shifted lobes respectively) which in part will be down to resolving out some extended emission, but in addition, we now partially resolve the outflow, clarifying what is and what is not associated with it. The ACA should be sensitive to $\sim$30$''$ scales at these wavelengths, although the \emph{u,v} coverage and short observations ($\sim$3\,min) mean there are negative artefacts in the images and we did not mosaic the region. Figure \ref{fig:fig11} is the channel map plot for the $^{13}$CO outflow emission. For the blue-shifted lobe, for velocities between 16.9 to 20.0\,km\,s$^{-1}$ the emission is unresolved in the $\sim$7$\times$5$''$ beam and located slightly south-east to the peak of the continuum emission. There is no emission near G17.64 from 20.0 to 21.0\,km\,s$^{-1}$ (possible self-absorbed or foreground absorption), while the $^{13}$CO forms a wide V-like shape cavity in the south-east from 21.5 to 23.0\,km\,s$^{-1}$. In conjunction with the main array observations this lower-velocity $^{13}$CO emission appears most fitting with the blue-shifted cavity. Such a shape would be expected if we are seeing the rear wall of the blue-shifted outflow cavity which would be orientated more in the plane of the sky. If we include this emission in the mass calculation we obtain an estimate closer to $\sim$10\,M$_{\odot}$ (between 16.9 to 23.0\,km\,s$^{-1}$). There is weak emission at $\sim$25.0\,km\,s$^{-1}$ and the redshifted and spatially offset lobe begins from about 26.0\,km\,s$^{-1}$ and extends to 35.6\,km\,s$^{-1}$ in velocity and to the north-west in position. We note that in Fig.\,\ref{fig:fig11} the red-shifted emission past 30.5\,km\,s$^{-1}$ is not shown as it has the same structure from 30.5\,km\,s$^{-1}$ to 35.6\,km\,s$^{-1}$ 

\begin{figure*}[ht!]
\begin{center}
  \includegraphics[width=0.7\textwidth]{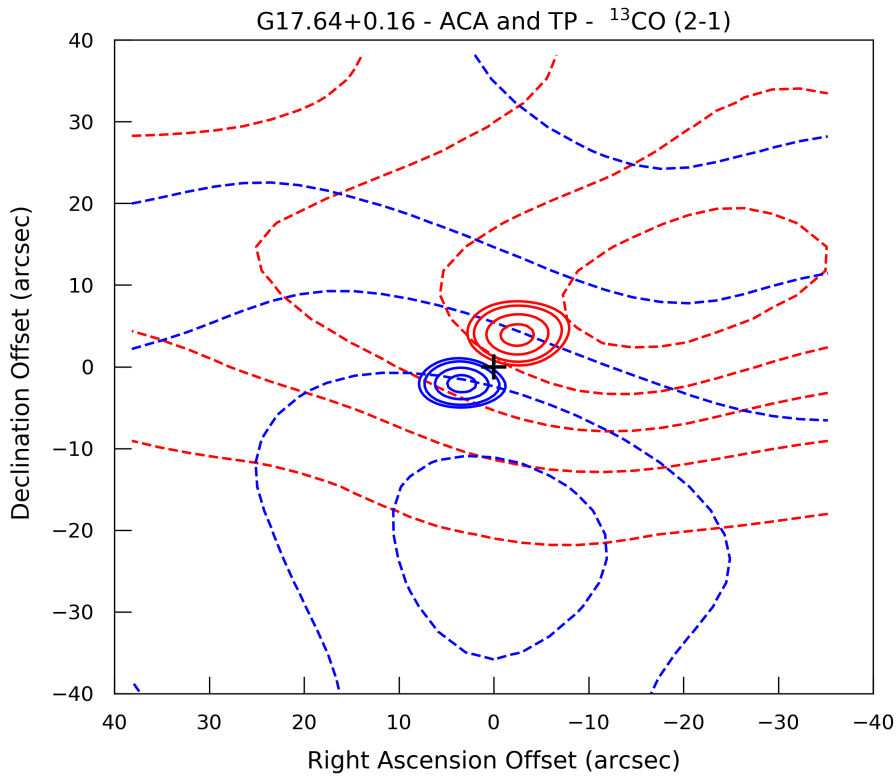}
  \caption{Contour plot of the $^{13}$CO outflow emission from the ACA (solid) and TP (dashed) observations. The blue- and red-shifted velocities are 18.5 to 20.1\,km\,s$^{-1}$ and 28.5 to 30.5\,km\,s$^{-1}$ respectively. The ACA contours are at 40, 50, 70 and 90\,\% of the peak value, while the total power are at 10, 30, 50, 70 and 90\,\% of the peak value. The plot is illustrative of the blue and red shift offset emission that forms the outflow, but also shows that the TP does not map a large enough region to fully map the entire CO emission.}
\label{fig:fig12} 
\end{center}
\end{figure*}

\begin{figure*}
\begin{center}
  \includegraphics[width=0.95\textwidth]{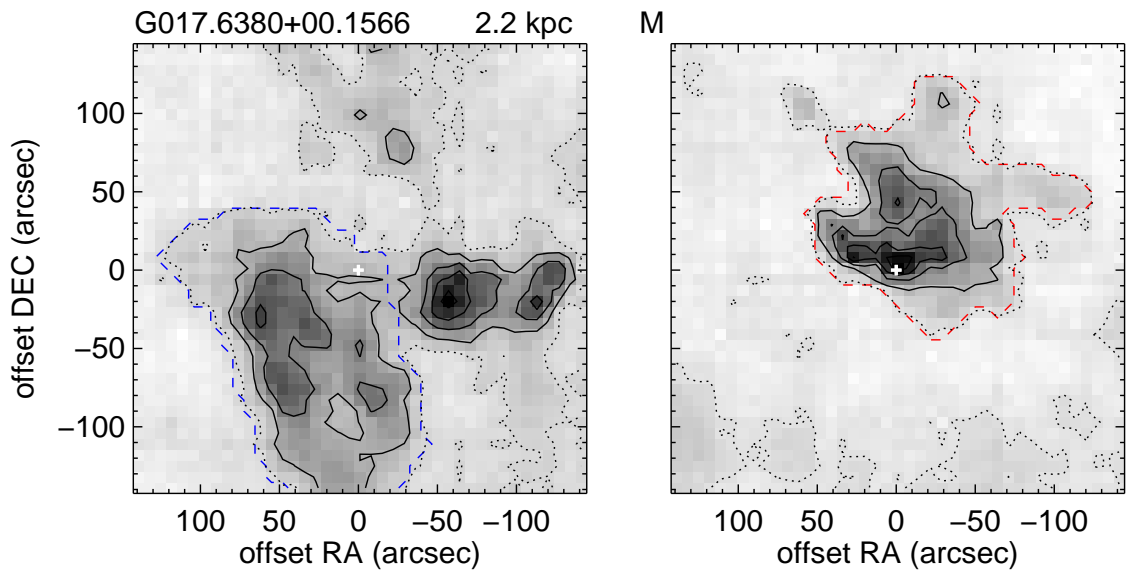}
\caption{Maps of the blue- (left) and red-shifted (right) $^{12}$CO (3-2) emission analysed as part of the JCMT outflow survey conducted by \citet{Maud2015}. The integrated ranges for the blue- and red-shifted emission are 6.0 to 20.0\,km\,s$^{-1}$ and 25.9 to 38.2\,km\,s$^{-1}$. These maps indicate the size of the emission thought to be associated with the outflow as compared with our ALMA ACA and TP data in Fig.\,\ref{fig:fig12}.}
\label{fig:fig13} 
\end{center}
\end{figure*}

In general we agree with previous studies indicating the orientation of the outflow: it appears blue-shifted in the south-east and is red-shifted to the north-west with a PA consistent with 135$^{\circ}$ \citep{Kastner1994,Maud2015}. However, between the ACA observations and the disc focussed main 12\,m array observations, we either do not have enough resolution to disentangle the morphology (ACA), or at high resolution we resolve out too much extended structure (12\,m data). In Fig.\,\ref{fig:fig12} we show the blue- and red-shifted outflow emission as contours from the ACA (solid) and Total Power - TP (dashed). It is also clear that the ACA is not sensitive to these larger scales, although the TP emission does not map a large enough area to fully encompass all the outflow emission, as shown in Fig.\,\ref{fig:fig13} from the work by \citet{Maud2015}. Full characterisation of the outflow in the region would require a mosaic in total-power, ACA observations and at least one 12\,m compact ($<$1000\,m baselines) array configuration (observing the $^{12}$CO, $^{13}$CO and C$^{18}$O transitions) with high-sensitivity and an excellent \emph{u,v} coverage.

\end{document}